\documentclass[12pt]{aastex}
\usepackage{emulateapj5}

\newcommand{\um}{$\mu$m}
\newcommand{\kms}{\mbox{\,km\,s$^{-1}$}}
\newcommand{\Msun}{\,$M_{\odot}$}

\newcommand{\hii}{\mbox{$\mathrm{H\,{\scriptstyle {II}}}$}}

\newcommand{\tco}{\mbox{$^{13}${\rmfamily CO}}}

\newcommand{\MSX}{{\it MSX}}
\newcommand{\ISO}{{\it ISO}}

\newcommand{\spitzer}{{\it Spitzer}}
\newcommand{\vlsr}{$V_{LSR}$}

\newcommand{\cms}{${\rm cm}^{-2}$}
\newcommand{\cmc}{${\rm cm}^{-3}$}

\newcommand{\bufcrao}{Boston University--Five College Radio Astronomy Observatory}
\newcommand{\irdcone}{G028.37+00.07}
\newcommand{\irdcthree}{G035.39$-$00.33}
\newcommand{\irdceight}{G035.59-00.24}
\newcommand{\irdcnine}{G024.60+00.08}
\newcommand{\irdctwelve}{G028.08+00.07}
\newcommand{\irdcfifteen}{G015.31$-$00.16}
\newcommand{\irdceighteen}{G053.11+00.05}
\newcommand{\irdctwentysix}{G028.28$-$00.34}

\newcommand{\irdcthirty}{G028.53$-$00.25}

\newcommand{\irdcthirtynine}{G033.69$-$00.01}
\newcommand{\irdcfortythree}{G034.43+00.24}
\newcommand{\irdcfortynine}{G022.73+00.11}
\newcommand{\irdcfiftyone}{G024.33+00.11}

\shorttitle{IRDCs: precursors to star clusters}
\shortauthors{Rathborne et al.}

\begin{document}

\title{Infrared Dark Clouds: precursors to star clusters}

\author{J. M. Rathborne and J. M. Jackson}
\affil{Institute for Astrophysical Research, Boston University, Boston, MA 02215; rathborn@bu.edu, jackson@bu.edu}
\and
\author{R. Simon}
\affil{I.Physikalisches Institut, Universit\"at zu K\"oln, 50937 K\"oln, Germany; simonr@ph1.uni-koeln.de} 

\begin{abstract}
Infrared Dark Clouds (IRDCs) are dense molecular clouds seen as
extinction features against the bright mid-infrared Galactic
background. Millimeter continuum maps toward 38 IRDCs reveal extended
cold dust emission to be associated with each of the IRDCs. IRDCs
range in morphology from filamentary to compact and have masses of 120
to 16,000\,\Msun, with a median mass of $\sim$940\,\Msun. Each IRDC
contains at least one compact ($\leq$0.5~pc) dust core and most show
multiple cores. We find 140 cold millimeter cores unassociated with
\MSX\, 8\,\um\, emission. The core masses range from 10 to
2,100\,\Msun, with a median mass of $\sim$120\,\Msun. The slope of the
IRDC core mass spectrum ($\alpha$ $\sim$ 2.1 $\pm$ 0.4) is similar to
that of the stellar IMF. Assuming that each core will form a single
star, the majority of the cores will form OB stars. IRDC cores have
similar sizes, masses, and densities as hot cores associated with
individual, young high-mass stars, but they are much colder. We
therefore suggest that IRDC represent an earlier evolutionary phase in
high-mass star formation. In addition, because IRDCs contain many
compact cores, and have the same sizes and masses as molecular clumps
associated with young clusters, we suggest that IRDCs are the cold
precursors to star clusters. Indeed, an estimate of the star formation
rate within molecular clumps with similar properties to IRDCs ($\sim$
2\,\Msun\,yr$^{-1}$) is comparable to the global star formation rate
in the Galaxy, supporting the idea that all stars may form in such
clumps.
\end{abstract}


\section{Introduction}

Although high-mass stars ($>$6\,\Msun) have a profound effect on the
Galactic environment and are responsible for recycling and enriching
interstellar matter, their early evolutionary phases are not well
known. Because they evolve rapidly, high-mass stars have a short
lifetime. In addition, high-mass stars emit copious amounts of UV
photons, which quickly heat, ionize, and disrupt their natal molecular
clouds. Thus, direct observations of their earliest stages are
difficult.

Because high-mass stars invariably form in star clusters, to
understand high-mass star formation it is important to study the
earliest stages in the formation of star clusters. Molecular line,
infrared (IR), and millimeter continuum studies of young, embedded
clusters show that star clusters form from small ($\sim$0.5--1 pc),
massive ($\sim$100--1000\,\Msun), and dense
($\sim$10$^{4}$--10$^{6}$\,\cmc) molecular clumps within a giant
molecular cloud (see Table~\ref{properties};
\citealp{Lada03,Motte03}). Cluster-forming clumps have star-formation
efficiencies of $\sim$ 10--30\% and a highly fragmented sub-structure
\citep{Lada97}. Within these clumps, the densest, most compact
sub-structures, called ``cores'', give rise to individual stars.

The emerging picture of star formation begins with the fragmentation
of a molecular cloud (e.g. \citealp{Shu87}). Observational evidence
suggests that molecular clouds are fragmented on all size scales
\citep{Williams00}. The cloud eventually condenses into cold, 
gravitationally bound starless cores. As the core becomes centrally
concentrated and begins to collapse gravitationally, a protostar and
disk system is formed as material is accreted from its
surroundings. Soon thereafter, bipolar outflows and jets are formed as
a consequence of the accretion process.  The central star then emerges
from the molecular cloud to be an optically visible main-sequence
star.

For low-mass stars, the theory is well developed (e.g.\
\citealp{Shu91}), and all of these evolutionary stages are observed
(e.g.\ \citealp{Lada99}). The earliest stage in the formation of
low-mass stars are identified as Bok globules.  These are isolated,
well-defined patches of optical obscuration (A$_{v}$=1--25 mag) viewed
against background stars \citep{Bok47}. Because of their isolation and
uncomplicated environments, Bok globules have been studied extensively
and have provided an excellent observational perspective of the
initial conditions within low-mass star-forming molecular clouds.
Many studies of Bok globules have used optical and IR star counts to
trace their visual extinctions and, hence, have measured their column
densities, masses, and internal structure
(e.g. \citealp{Lada94,Alves01,Lombardi01}). Bok globules have small
sizes (0.1--2\,pc), characteristic masses of {\mbox{1--100\,\Msun}},
and a simple morphology (see Table~\ref{properties};
\citealp{Leung85,Clemens88,Bourke95}).

Within Bok globules, the dense, compact pre-cursors to the individual
protostars are found. These so-called ``pre-protostellar cores'' have
low temperatures ($\sim$ 10 K), compact sizes ($\sim$0.05 pc), low
masses (0.5--5\,\Msun), and high densities (10$^{5}$--10$^{6}$ \cmc;
see Table~\ref{properties};
e.g. \citealp{Meyers83,Ward-Thompson94}). Once the embedded protostar
begins to evolve toward the main-sequence, it emits strongly in the IR
(e.g.\ \citealp{Lada84,Adams87}) and reveals molecular outflows, jets,
and Herbig-Haro objects (e.g. \citealp{Snell80,Bally83,Fukui93}).

In contrast, the early evolution of high-mass stars is not as well
understood. Observationally, the earliest known, well-characterized
phase of high-mass star formation is associated with a special class
of molecular cores called ``hot cores.'' Hot cores are internally
heated (50--250 K), small ($<$0.1 pc), massive
($\sim$100--300\,\Msun), and dense (10$^{5}$--10$^{8}$\,\cmc; see
Table~\ref{properties}; \citealp{Garay99,Kurtz00,Churchwell02}). In
the later stage of the hot core phase, molecular outflows and maser
emission appear, both of which are signatures of an accretion disk
\citep{Garay99,Kurtz00}. This stage marks the appearance of
ultra-compact \hii\, regions, the small, dense regions of ionized gas
which surround the embedded high-mass main-sequence star.

These observed early evolutionary stages of high-mass stars represent
a phase {\it after} the central protostar has already formed. A
complete understanding of the formation of high-mass stars, however,
must begin with an earlier phase, before the formation of the
protostar. Presumably, high-mass stars will also proceed through
stages analogous to the Bok globules and pre-protostellar cores found
in the case of low-mass star formation. So far, however, observational
examples of these phases for high-mass stars have remained elusive.

By analogy, we expect the high-mass equivalents of Bok globules and
high-mass starless cores also to be cold. However, because they are
the precursors to high-mass stars, they should have larger sizes,
masses, column densities, and volume densities.  Because the high-mass
Bok globule analogs are likely to be cold and very dense, one might
expect to detect them as extinction features and to trace their
internal structure through millimeter/submillimeter continuum imaging.

We suggest that the earliest stage of high-mass star formation occurs
within infrared dark clouds (IRDCs). IRDCs are defined as regions of
high extinction viewed against the bright, diffuse mid-IR Galactic
background. IRDCs, first detected by \ISO\, and then by \MSX\,
\citep{Perault96,Egan98,Hennebelle01}, are ubiquitous throughout the
Galaxy \citep{Simon-catalog}. Previous studies show that their
molecular material has low temperatures ($<$ 25K), high column
densities ($\sim$ 10$^{23}$--10$^{25}$\,\cms), and high volume
densities ($>$10$^{5}$ \,\cmc; \citealp{Egan98,Carey98,Carey00}). A
recent molecular line study of a large sample of IRDCs establishes
their kinematic distances and shows that their Galactic distribution
is enhanced towards the Galaxy's most massive star forming structure,
the so-called 5\,kpc ring \citep{Simon-msxgrs}. Millimeter/submillimeter studies of a few IRDCs
show that they harbor compact cores
\citep{Lis94,Carey00,Redman03,Garay04,Ormel05,Rathborne05}.

We have developed an algorithm \citep{Simon-catalog} to identify IRDCs
as regions within the \MSX\, 8\,\um\, Galactic Plane Survey
\citep{Price01} that have a significant decremental
contrast\footnote{Defined as contrast=(background-image)/background.} 
against a modelled background. IRDCs were identified from contrast
images if they met two conditions; (1) they must be regions of
contiguous pixels whose measured contrast lies at least 2$\sigma$
above the instrumental noise, and (2) they must have solid angles $>$
1,200 square arcsec. Because the Galactic background is highly
variable, the signal-to-noise, and hence the significance level, is
non-uniform for this sample.
\cite{Simon-catalog} find 10,961 IRDCs meeting
these criteria\footnote{IRDCs from this catalog are called ``MSXDC G'' followed by their
Galactic coordinates. For simplicity we will drop the ``MSXDC'' labels in the text.}.

Because IRDCs are extinction features seen against the Galactic mid-IR
background, their identification is strongly biased toward nearby
clouds and clouds that lie in front of bright mid-IR background
emission.  Almost certainly, clouds with identical temperatures,
sizes, masses, column densities, and volume densities as known IRDCs
have escaped detection. These undetected clouds will preferentially
lie on the far side of the Galaxy, behind the bulk of the Galaxy's
diffuse mid-IR emission, or in the outer Galaxy, where the Galactic
mid-IR background is faint.

Because IRDCs are cold, their thermal dust emission peaks at
millimeter/submillimeter wavelengths.  Consequently, we can best study
their internal structure by imaging at these wavelengths.  A further
advantage is that the dust emission at millimeter wavelengths is
optically thin. Thus, unlike optically thick molecular line emission,
the millimeter dust continuum is a more accurate tracer of column
density and mass.

In this paper, we show that IRDCs are cold molecular clumps with sizes
and masses identical to those of the warmer cluster-forming
clumps. Moreover, we find that IRDCs contain compact cores with sizes,
masses, and densities comparable to high-mass star-forming hot cores,
the only obvious difference being that IRDC cores are colder. We
therefore suggest that IRDCs are the cluster-forming, high-mass
analogs to Bok globules, and their embedded cores the precursors to
individual high-mass protostars.

\section{Observations}

For this study, we selected the 38 darkest IRDCs with known kinematic distances from the sample of 
\cite{Simon-msxgrs}. The local standard of rest (LSR) velocity for each IRDC was determined 
from a morphological match of the mid-IR extinction to \tco\, emission
from the \bufcrao\, Galactic Ring Survey
\citep{Simon01,Jackson06}. Velocities were converted to a
Galactocentric radius and kinematic distance using the
\cite{Clemens85} rotation curve\footnote{For the portion of the Galaxy
sampled, the use of the more recent \cite{Brand93} rotation curve does
not significantly alter the derived distances and therefore will not
affect any of our conclusions.}  scaled to R$_{0}$=8.5 kpc and
$\Theta_{0}$=220\,\kms. Because IRDCs are extinction features, they
are assumed to lie at the near kinematic distance. The IRDCs within
our sample show the highest peak contrast relative to the extended
mid-IR background ($>$ 36\%) and span a broad range in distance and
Galactic location.

The 1.2\,mm continuum data toward the IRDCs were obtained at the
Institut de Radioastronomie Millim\'etrique (IRAM) 30\,m telescope
using the 117 element bolometer array MAMBO~II in 2003 December and
2004 January--March. The FWHM angular resolution of each element in
the array is 11\arcsec. The angular separation between array elements
is 20\arcsec. Because the morphology of the IRDCs ranged from compact
to filamentary, we mapped each of the IRDCs using the on-the-fly
mapping mode with a size large enough to cover the extent of the
mid-IR extinction. The map sizes ranged from 3$\arcmin \times
3\arcmin$ to 9$\arcmin \times 9\arcmin$.  The scanning speed was 6
arcsec s$^{-1}$ for maps 6$\arcmin \times 6\arcmin$ or smaller, and
8~arcsec~s$^{-1}$ for larger maps. To sample the emission fully, all
maps were obtained in the `sweeping' mode, where the spacing between
sub-scans was set to 22$\arcsec$. Pointing and sky-dip calibrations
were performed regularly.

All data reduction was achieved within the MOPSI
package\footnote{MOPSI was developed by Robert Zylka.}.  All maps were
reduced by applying the atmospheric opacity corrections, fitting and
subtracting a baseline, and removing the correlated sky noise. Maps
were flux calibrated using the conversion factor obtained from the
counts detected in observations of Uranus. The rms noise level in each
of the final maps is $\sim$ 10 mJy beam$^{-1}$.

\section{Results}

\subsection{Global Millimeter Emission Characteristics}

\begin{figure*}
\includegraphics[width=0.75\textwidth]{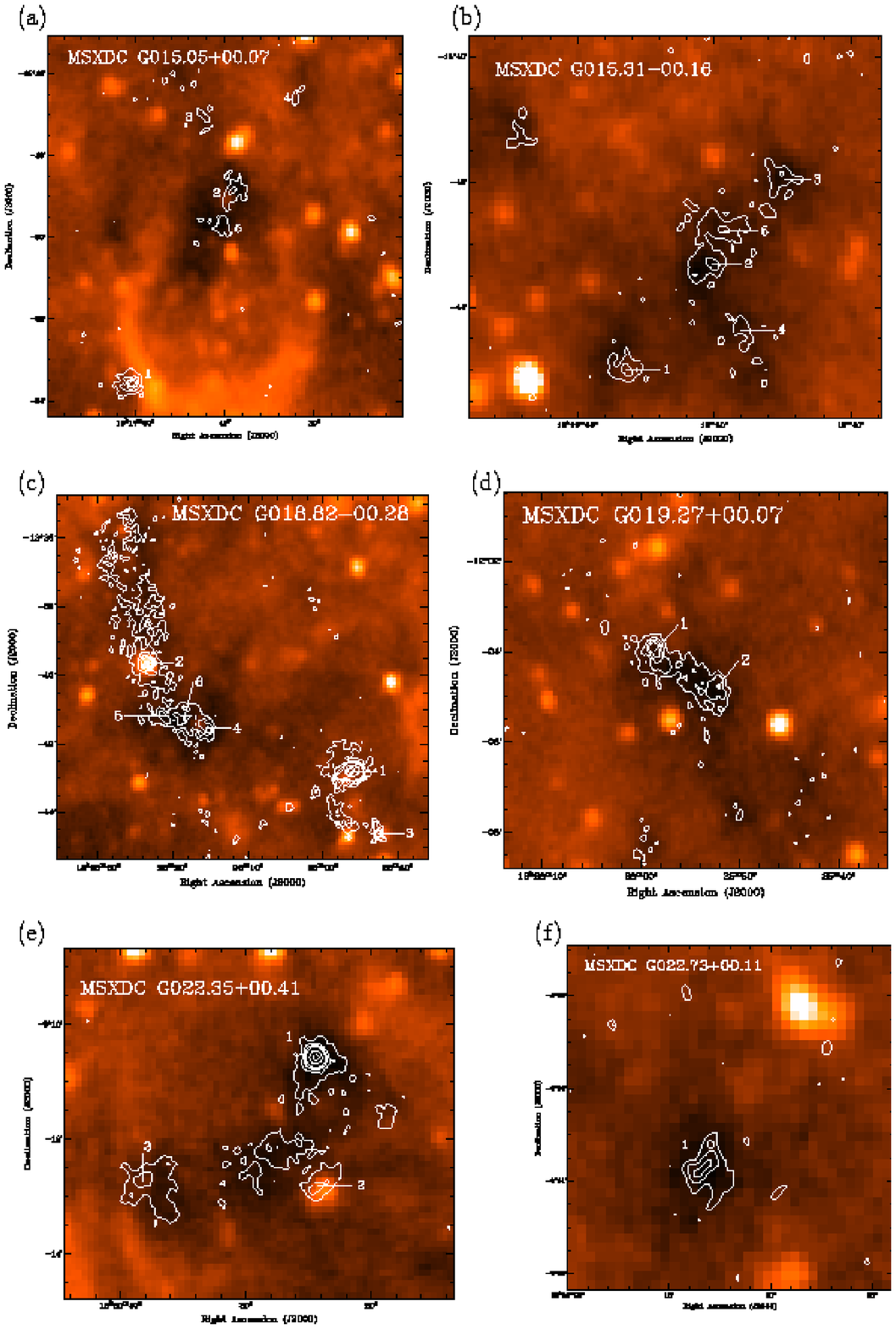}\\
\caption{\label{mm-fig1} \MSX\, 8\,\um\, images overlaid with 1.2\,mm continuum emission for six IRDCs. The 1.2 mm
cores identified within the IRDCs are labeled with their designation (as listed in Table~\ref{table-cores}). The contours
in all cases are 30 ($\sim$3$\sigma$), 60, 90, 120, 240, 360, 480, 840, 1200, 2400\,mJy beam$^{-1}$, except for 
\irdcfifteen\, which has contours of 20 and 40 mJy beam $^{-1}$ and \irdcfortynine\,
which has contours of 20, 30, 45\,mJy beam$^{-1}$.}
\end{figure*}

\begin{figure*}
\includegraphics[width=0.75\textwidth]{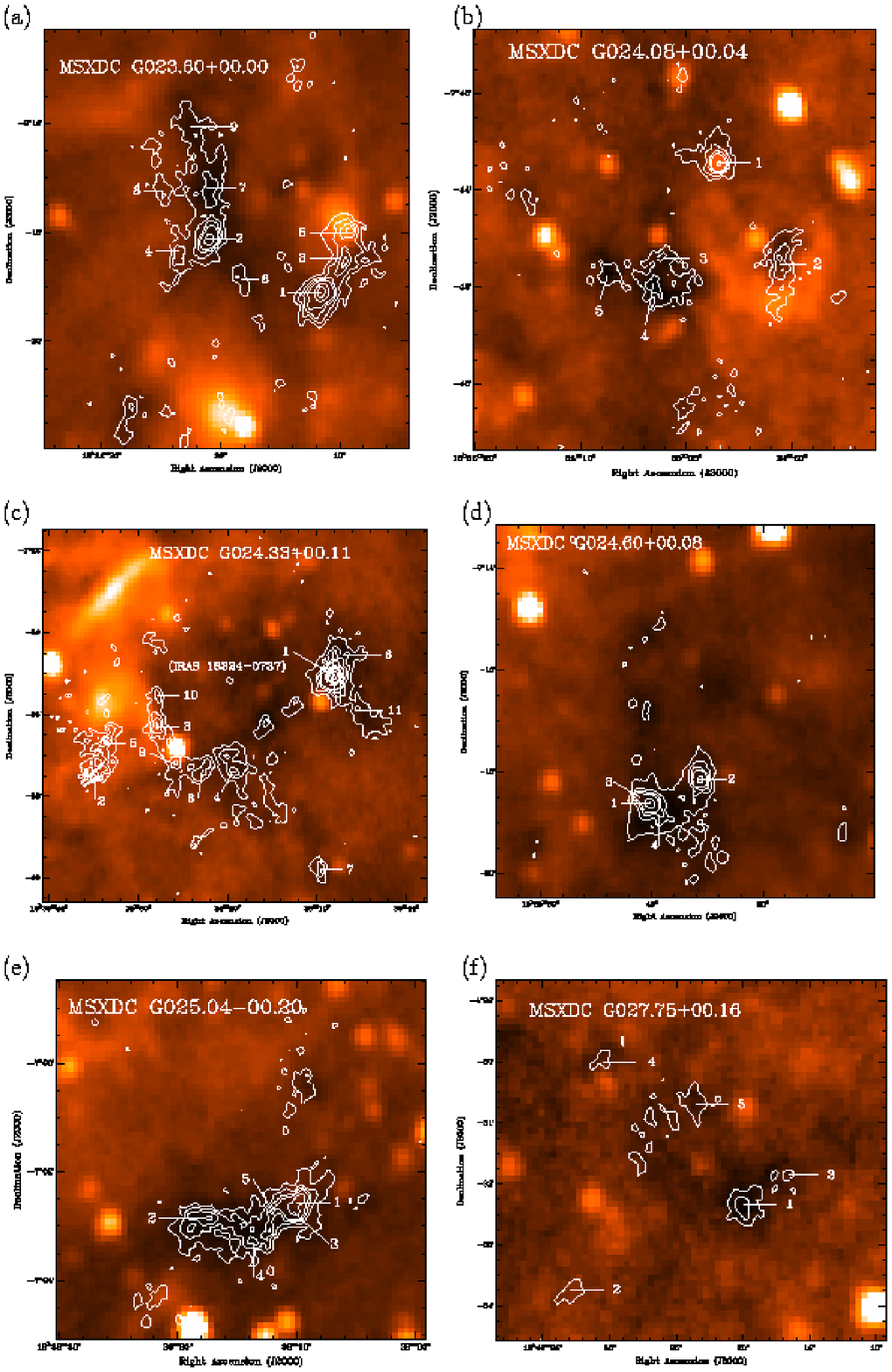}\\
\caption{\label{mm-fig2} \MSX\, 8\,\um\, images overlaid with 1.2\,mm continuum emission for six IRDCs. The 1.2 mm
cores identified within the IRDCs are labeled with their designation (as listed in Table~\ref{table-cores}). The contours
in all cases are 30 ($\sim$3$\sigma$), 60, 90, 120, 240, 360, 480, 840, 1200, 2400\,mJy beam$^{-1}$. The ellipse overlaid
on \irdcfiftyone\, marks the position of the IRAS source.}
\end{figure*}

\begin{figure*}
\includegraphics[width=0.75\textwidth]{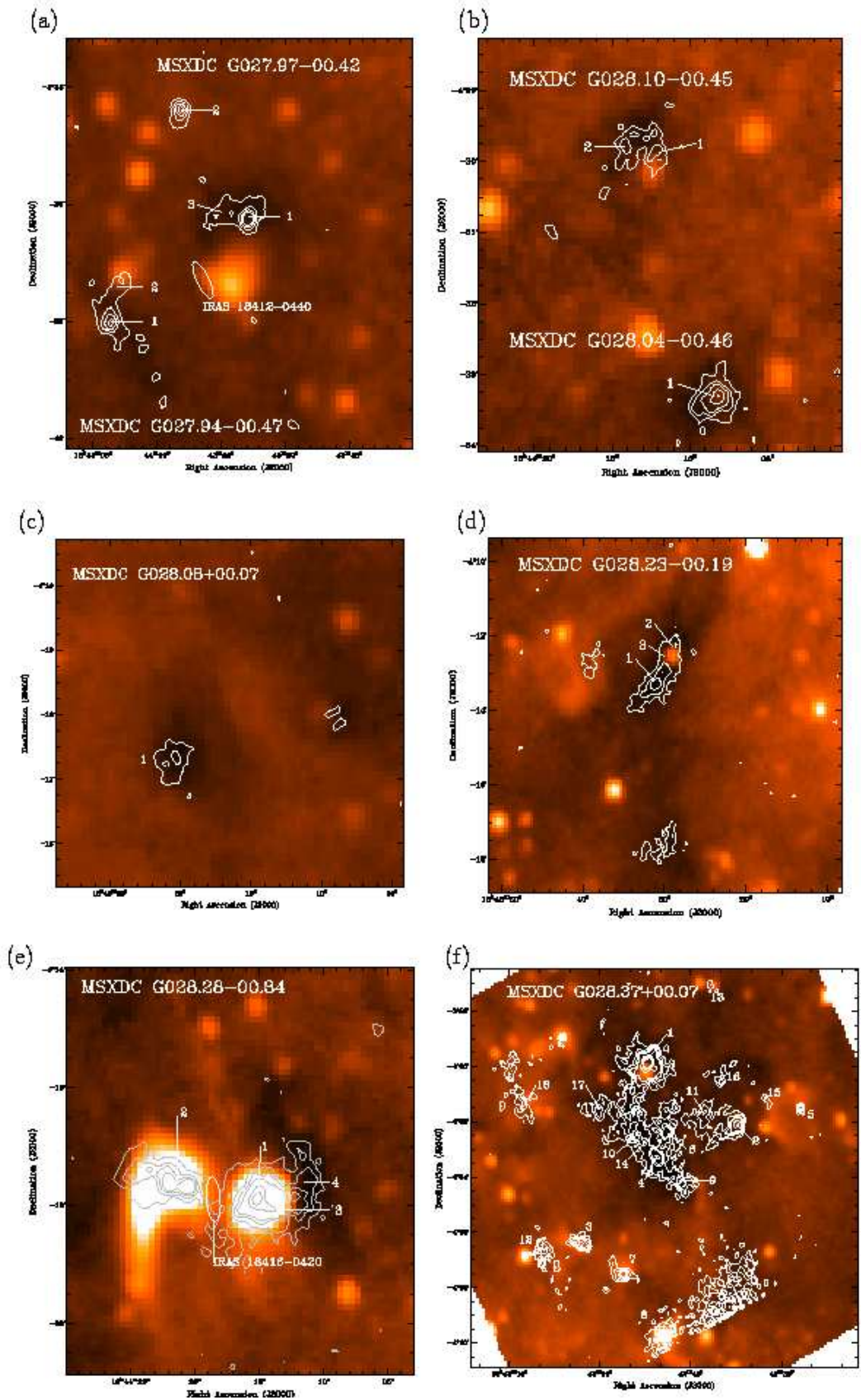}\\
\caption{\label{mm-fig3}\MSX\, 8\,\um\, images overlaid with 1.2\,mm continuum emission for six IRDCs. The 1.2 mm
cores identified within the IRDCs are labeled with their designation (as listed in Table~\ref{table-cores}). The contours
in all cases are 30 ($\sim$3$\sigma$), 60, 90, 120, 240, 360, 480, 840, 1200, 2400\,mJy beam$^{-1}$. The ellipses mark
the position of the labeled IRAS sources.}
\end{figure*}

\begin{figure*}
\includegraphics[width=0.75\textwidth]{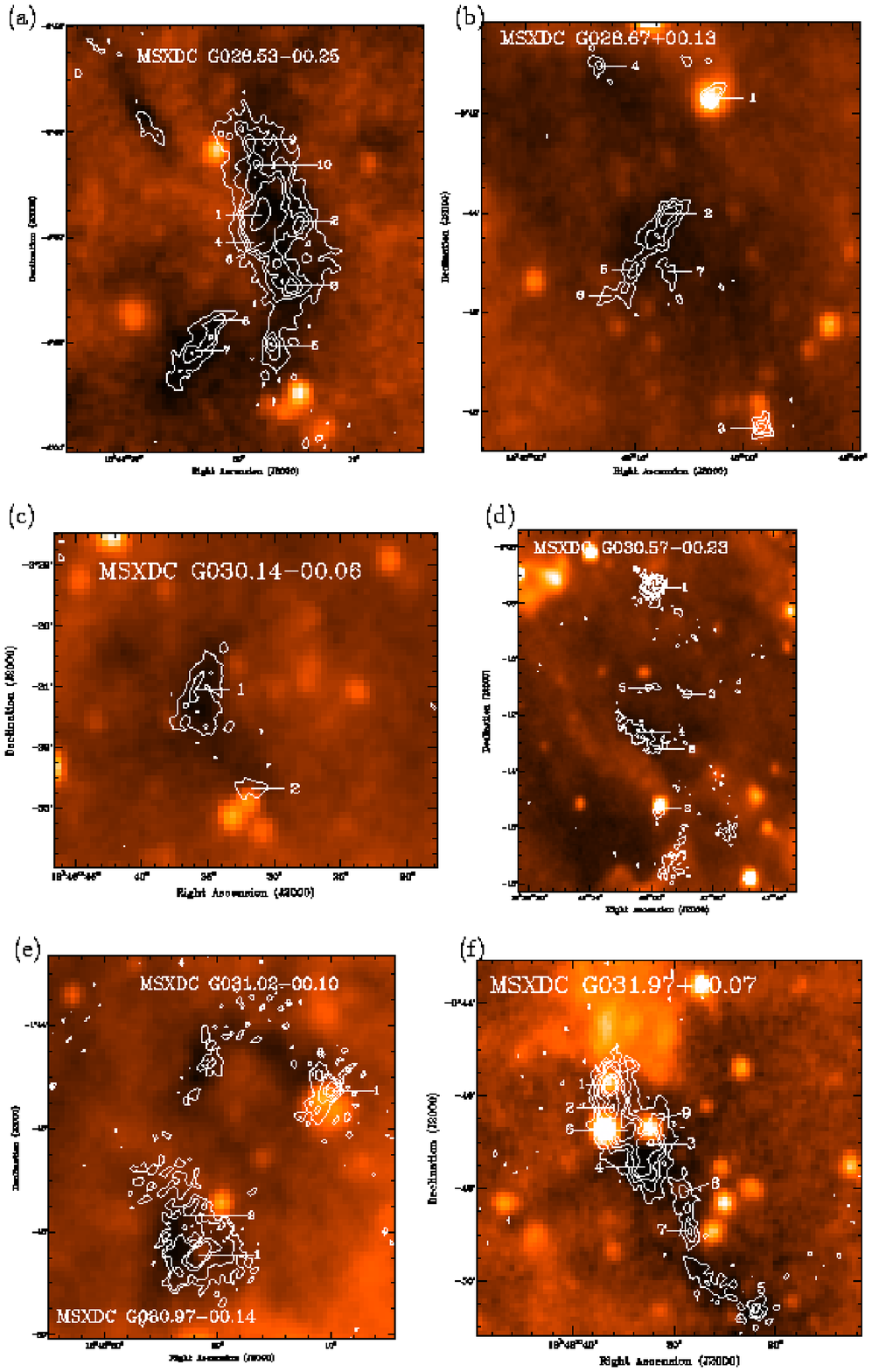}\\
\caption{\label{mm-fig4}\MSX\, 8\,\um\, images overlaid with 1.2\,mm continuum emission for six IRDCs. The 1.2 mm
cores identified within the IRDCs are labeled with their designation (as listed in Table~\ref{table-cores}). The contours
in all cases are 30 ($\sim$3$\sigma$), 60, 90, 120, 240, 360, 480, 840, 1200, 2400\,mJy beam$^{-1}$.}
\end{figure*}

\begin{figure*}
\includegraphics[width=0.75\textwidth]{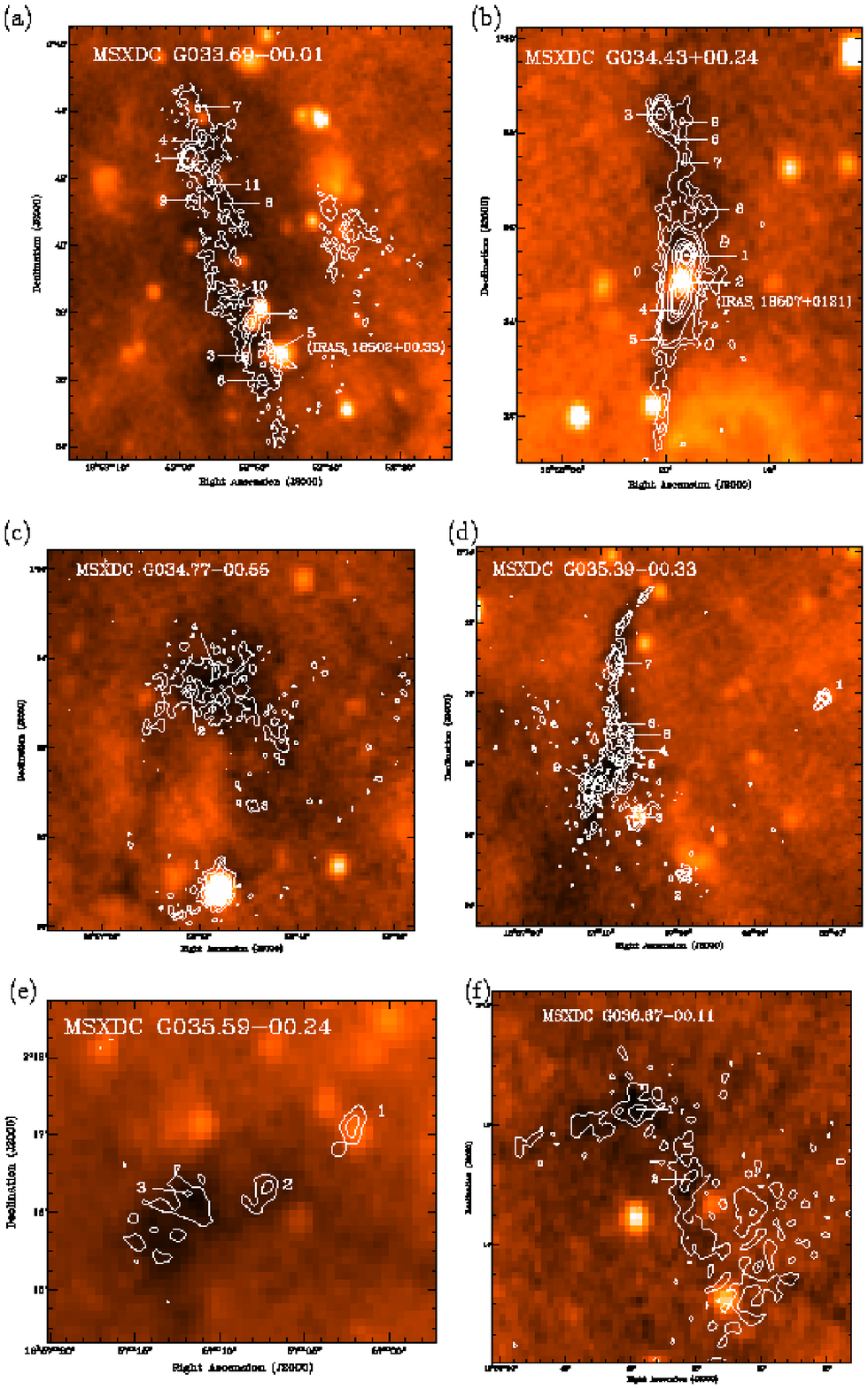}\\
\caption{\label{mm-fig5}\MSX\, 8\,\um\, images overlaid with 1.2\,mm continuum emission for six IRDCs. The 1.2 mm
cores identified within the IRDCs are labeled with their designation (as listed in Table~\ref{table-cores}). The contours
in all cases are 30 ($\sim$3$\sigma$), 60, 90, 120, 240, 360, 480, 840, 1200, 2400\,mJy beam$^{-1}$, except for
\irdcfortythree\, which starts at 60 mJy beam$^{-1}$. The ellipse marks the position of the IRAS source.}
\end{figure*}

\begin{figure*}
\includegraphics[width=0.75\textwidth]{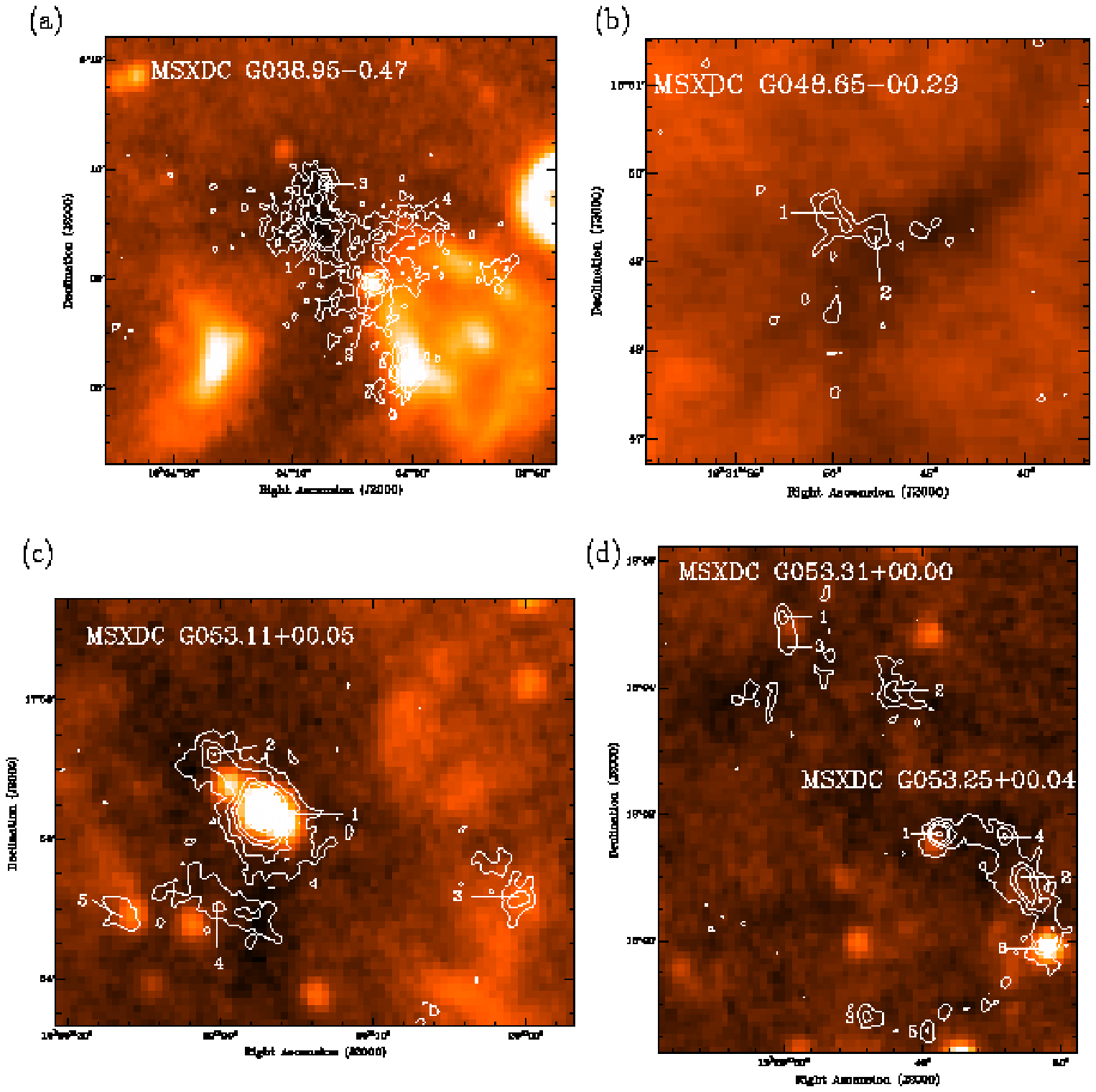}\\
\caption{\label{mm-fig6}\MSX\, 8\,\um\, images overlaid with 1.2\,mm continuum emission for six IRDCs. The 1.2 mm
cores identified within the IRDCs are labeled with their designation (as listed in Table~\ref{table-cores}). The contours
in all cases are 30 ($\sim$3$\sigma$), 60, 90, 120, 240, 360, 480, 840, 1200, 2400\,mJy beam$^{-1}$.}
\end{figure*}

Millimeter continuum emission was detected toward all 38 of the IRDCs
observed. Figures~\ref{mm-fig1}--\ref{mm-fig6} show the 1.2\,mm
continuum emission toward each IRDC overlaid on the \MSX\, 8\,\um\,
image. In all cases, the morphology of the millimeter continuum
emission matches the mid-IR extinction extremely well.In addition, all
IRDCs have at least one bright, compact millimeter source; most show
multiple, compact sources superimposed on fainter, extended emission.

Although IRDCs all exhibit extinction in the mid-IR, their
morphologies vary considerably.  Their shapes can be simple and
unstructured on the one hand, and complex and filamentary on the
other. For example, the IRDCs \irdctwelve\, (Fig.~\ref{mm-fig3}[c])
and \irdceight\, (Fig.~\ref{mm-fig5}[e]) show little sub-structure,
whereas \irdcfortythree\, (Fig.~\ref{mm-fig5}[b]) and \irdcthree\,
(Fig.~\ref{mm-fig5}[d]) are highly filamentary. Some IRDCs also show a
complicated internal sub-structure with 10 or more compact cores, for
example, \irdcfiftyone\, (Fig.~\ref{mm-fig2}[c]), \irdcone\,
(Fig.~\ref{mm-fig3}[f]), \irdcthirty\, (Fig.~\ref{mm-fig4}[a]), and
\irdcthirtynine\, (Fig.~\ref{mm-fig5}[a]).

By eye we have catergorized each IRDC as either ``compact'' or
``filamentary'' based on the morphology of the millimeter continuum
emission. We find 22 IRDCs are compact, while 16 are filamentary. We
find a slight trend for higher mass cores to be found within the
filamentary clouds. For compact IRDCs the average mass of the most
massive core is $\sim$~210\,\Msun. For filamentary IRDCs, however, the
average mass of the most massive core is $\sim$~660\,\Msun, about a
factor of 3 higher.

The degree to which IRDCs are associated with mid-IR continuum
emission sources is highly variable. For example, \irdcfifteen\,
(Fig.~\ref{mm-fig1}[b]), \irdcfortynine\, (Fig.~\ref{mm-fig1}[f]), and
\irdcnine\, (Fig.~\ref{mm-fig2}[d]) show no obvious association with
any mid-IR continuum emission source in \MSX\, 8\,\um\, images (to a
sensitivity limit of 1.3~MJy sr$^{-1}$; \citealp{Price01}). On the
other hand, \irdctwentysix\, (Fig.~\ref{mm-fig3}[e]) and
\irdceighteen\, (Fig.~\ref{mm-fig6}[c]) appear to be associated with
very bright, resolved mid-IR emission sources. There is no obvious
correlation between the presence of a bright mid-IR emission source
with any millimeter continuum observational characteristic of the
IRDC.

\subsection{IRDC Masses}

Reliable mass estimates of IRDCs using molecular lines are problematic
because the lines are often highly opaque and thus do not trace their
embedded interiors. In addition, because IRDCs are so cold, molecules
are frozen onto dust grains in their centers. Because dust is
optically thin and not depleted, it is a better mass tracer. We use
the 1.2~mm continuum emission to estimate the cloud mass using the
expression
\begin{equation}
M = \frac{F_{\nu} D^{2}}{\kappa_{\nu} B_{\nu} (T)}
\end{equation}
\noindent \citep{Hildebrand83}, where $F_{\nu}$ is the observed integrated source flux density, 
$D$ is the distance, $\kappa_{\nu}$ is the dust opacity per gram of
dust, and $B_{\nu}(T)$ is the Planck function at the dust
temperature. We adopt for $\kappa_{1.2 mm}$ a value of 1.0~cm$^2$
g$^{-1}$
\citep{Ossenkopf94} and assume a gas-to-dust mass ratio of 100. We assume graybody emission
with an emissivity index $\beta$ of 2 \citep{Carey00,Rathborne05}. We
also assume a dust temperature of 15 K, which is consistent with
temperature estimates from molecular line emission \citep{Teyssier02},
previous observations of a few IRDCs at submillimeter wavelengths
\citep{Carey00}, and the fact that some IRDCs are seen as extinction features up to 
100\,\um\, \citep{Egan98}.

The derived masses will be underestimated if the value for $\beta$ is
lower than the assumed value of 2 (e.g., $\beta$ of 1 will increase
the IRDC masses by about 10\%) and overestimated if the temperature is
higher than the assumed value of 15 K (e.g., a dust temperature of 30
K will reduce the IRDC masses by $\sim$ 40\%). The derived masses are
proportional to the assumed value of $\kappa_{\nu}$.

Table~\ref{table-clouds} lists, in order of increasing Galactic
longitude, each of the IRDCs, the coordinates of the peak contrast
\citep{Simon-catalog}, the LSR velocity (\vlsr), line width ($\Delta
V$) and kinematic distance ($D$; \citealp{Simon-msxgrs}), the peak and
integrated 1.2 mm flux, and the mass. The integrated 1.2 mm fluxes
were obtained by summing the 1.2 mm continuum emission above a
3$\sigma$ level ($>$30 mJy beam$^{-1}$) over the extent of the mid-IR
extinction feature. The total mass of the IRDCs were calculated using
the integrated flux and were found to range from 120 to 16,000\,\Msun,
with a median mass of 940\,\Msun\, and an average mass of $\sim$
3,000\,\Msun. Figures~\ref{mm-fig1}--\ref{mm-fig6} show that the
millimeter emission associated with each IRDC typically extends a few
arcmins, or a few parsecs given their distances.

\subsection{Compact Cores}

Compact cores were identified by fitting two-dimensional gaussians to
the peaks ($>$8$\sigma$) within the 1.2~mm continuum images. Each
Gaussian was subtracted before the next was fit until the image
contained only extended emission. The number of cores found within
each IRDC ranged from 1 to 18. Toward the 38 IRDCs, we detected a
total of 188 compact cores, increasing the sample of known cores
within IRDCs from $\sim$10 \citep{Carey00,Redman03,Garay04}.

Table~\ref{table-cores} lists the properties of the cores, including:
their designation, coordinates, peak and integrated 1.2 mm flux,
average angular and physical sizes, and mass. The cores are named in
order of decreasing peak 1.2 mm flux with the designation ``MM''
(e.g., MM1, MM2, etc). Because the 1.2~mm images extend beyond the
IRDC, they often include nearby, bright millimeter cores that are
associated with strong 8\,\um\, emission sources (and in some cases
IRAS sources). We find 48 of the cores are associated with
\MSX\, 8\,\um\, emission. These ``warm'' cores are included within our list, but are marked as being associated 
with 8\,\um\, emission and not an extinction feature (an ``em'' in the
comment column of Table~\ref{table-cores}). These cores are not
included in the following analyses. Thus, removing these ``warm''
cores from our sample, we find 140 ``cold'' millimeter cores
associated with the 38 IRDCs.

\begin{figure}
\begin{centering}
\epsscale{0.5}
\plotone{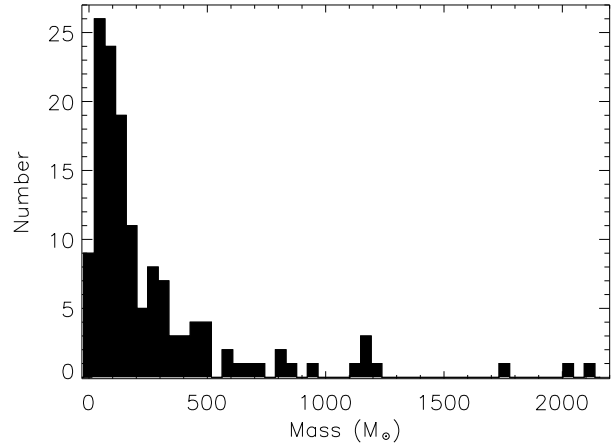}
\caption{\label{histogram} Number distribution of compact 1.2 mm cores with mass. We find the 
cores have masses in the range 10 to 2100\,\Msun\, with a median mass of 120\,\Msun. 
A power law fit to the curve (above a mass of 100\,\Msun) gives a slope of 2.1$\pm$0.4.
We find that $\sim$ 67\% of our sample lies in the mass range 30 to 300\,\Msun.  The
apparent peak near M $\sim$45\,\Msun\, reflects our detection limit.}
\end{centering}
\end{figure}

Masses for these compact 1.2 mm cores were calculated from the
integrated flux from the two-dimensional Gaussian fit to the core. A
histogram of the derived masses for the ``cold'' compact cores is
shown in Figure~\ref{histogram}. These cores range in mass from 10 to
2100\,\Msun, with a median value of 120\,\Msun.  We find that 67\% of
our sample lies in the mass range 30 to 300\,\Msun.  For masses
$>$100\,\Msun\, the distribution is roughly a power law with an index
$\alpha$ ($d$N/$d$M $\sim$ M$^{-\alpha}$) of 2.1 $\pm$ 0.4. The
apparent peak near M$\sim$45\,\Msun\, reflects our completeness limit.
Because the cores lie at different distances, the mass detection limit
varies from cloud to cloud. At the median distance of our sample of
$\sim$4.0~kpc, the 8$\sigma$ core detection limit corresponds to a
mass of $\sim$30~\Msun. Hence, below $\sim$ 45\,\Msun, our ``cold''
core sample is incomplete.

For three active star-forming cores within \irdcfortythree,
\cite{Rathborne05} find that greybody fits to their IR to millimeter
continuum spectral energy distributions reveal temperatures of $\sim$
30K. Thus, these masses will be overestimates if any of the cores have
a higher temperature than 15\,K (e.g., a dust temperature of 30\,K
will reduce the core mass by $\sim$ 40\%).

\begin{figure}
\begin{centering}
\epsscale{0.5}
\plotone{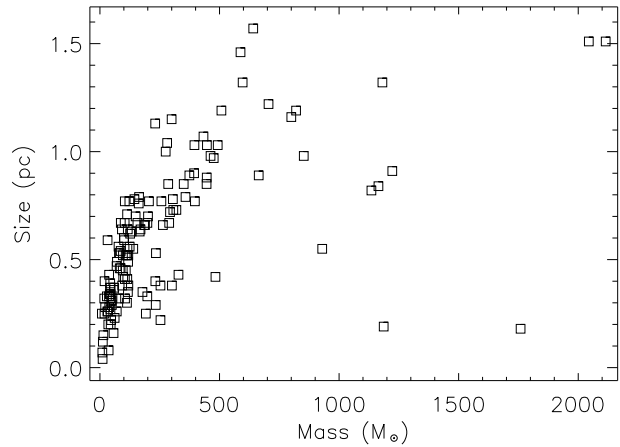}
\caption{\label{mass-size}Size versus mass for the compact 1.2 mm cores. The cores have sizes in the 
range 0.04 to 1.6 pc, with a median value of 0.5 pc.}
\end{centering}
\end{figure}

\begin{figure}
\begin{centering}
\epsscale{1.0}
\plotone{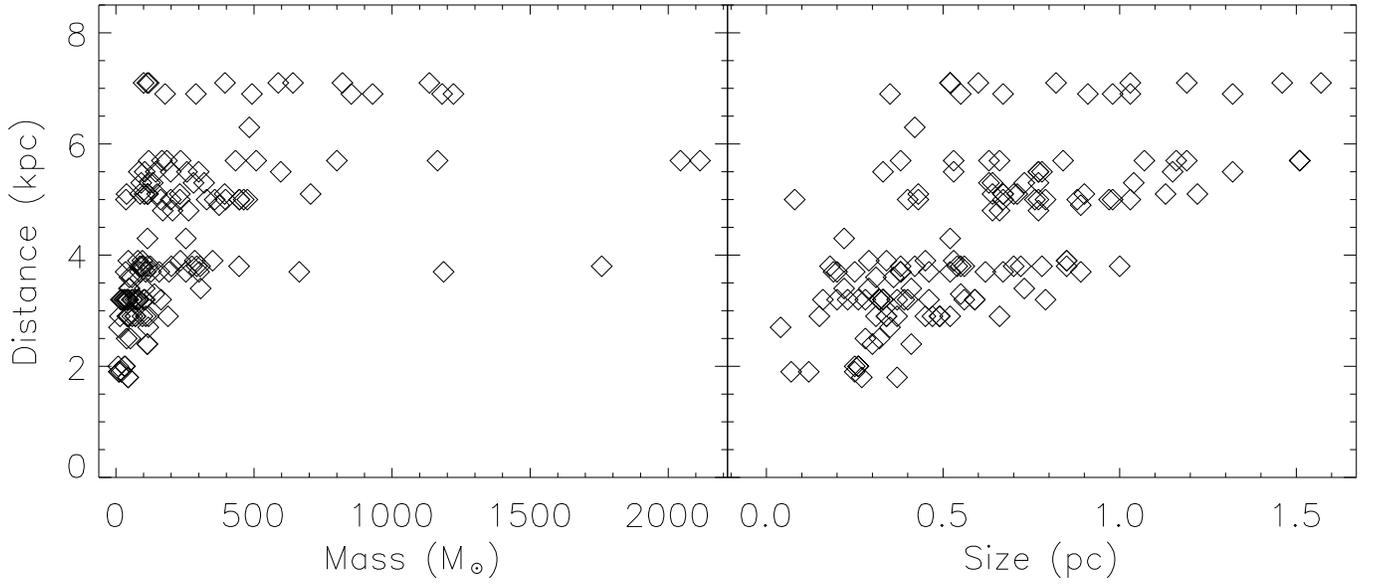}
\caption{\label{dist-mass-size}Distance as a function of 1.2 mm core mass (left) and size (right). We find no
correlation between either mass or size and distance to the cores. Indeed, we detect low-mass cores
at all distances.}
\end{centering}
\end{figure}

The cores have diameters in the range 0.04 to 1.6 pc, with a median
diameter of 0.5 pc. All cores are resolved in the 11\arcsec\, IRAM
beam. Figure~\ref{mass-size} plots the size versus mass for the
millimeter cores. The majority of cores have masses $<$200\,\Msun\,
and sizes $<$0.8\,pc.  We find no correlation between either the mass
or size and kinematic distance (Fig.~\ref{dist-mass-size}). Indeed, we
detect low-mass cores at all distances.

\begin{figure}
\begin{centering}
\epsscale{0.5}
\plotone{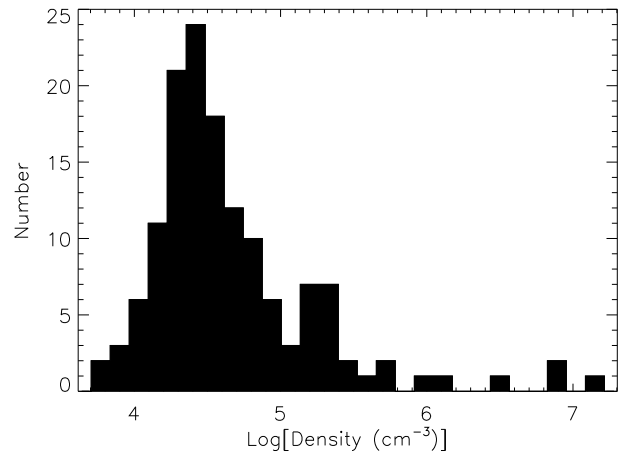}
\caption{\label{density} Number distribution of compact 1.2 mm cores with density. We find the densities
range from $10^{3}$ to $10^{7}$ \cmc, with a median value of 3.3 $\times$ 10$^{4}$ \cmc.}
\end{centering}
\end{figure}

Figure~\ref{density} plots the histogram of the core densities. The densities were
calculated using 

\begin{equation}
\rho =  \frac{3 M}{4 \pi r^{3}}
\end{equation}

\noindent where $M$ is the 1.2 mm mass and $r$ is the radius (half the size given 
in Table~\ref{table-cores}). 
The densities range from 10$^{3}$ to 10$^{7}$\,\cmc,
with a median value of 3.3 $\times$ 10$^{4}$\,\cmc.

\section{Discussion}

The combination of low optical depth and good angular resolution of
these millimeter images allow us to probe the internal structure of
IRDCs and to elucidate their role in the star formation process.

The sizes and masses of IRDCs as a whole are larger than those of Bok
globules and hot cores, but smaller than those of giant molecular
clouds. Instead, IRDCs more closely resemble cluster-forming molecular
clumps. Indeed, IRDC sizes ($\sim$ 2 pc) and masses
($\sim$10$^{2}$--10$^{4}$\,\Msun) are comparable to those of
cluster-forming clumps (see Table~\ref{properties}). Moreover they
both exhibit a highly fragmented sub-structure. A significant
difference, however, is that IRDCs are much colder.

\subsection{A Proposed Evolutionary Sequence for IRDCs}

Because the masses and sizes of IRDCs and cluster forming clumps are
so similar, we suggest that IRDCs are the high-mass analogs to Bok
globules.  As such, they are the precursors to star clusters. Their
lower temperatures would then result purely from an evolutionary
effect. IRDCs may be colder because they represent an early stage of a
cluster-forming molecular clump before the stars have formed and
heated the surrounding gas and dust.

If this scenario is correct, clusters begin their lives as cold,
pristine IRDCs (in this discussion, we will refer to {\it all}
molecular clumps with similar temperatures [$\sim$ 15 K], sizes
[$\sim$ 2 pc], masses [$\sim$10$^{2}$--10$^{4}$\,\Msun], and densities
[$\sim$10$^{2}$--10$^{4}$\,\cmc] as the \MSX\, dark clouds as
``IRDCs'', whether or not they are actually observed as extinction
features). As cores begin to collapse within the clump, protostars
form and begin to heat their local environment. If high-mass
protostars are present, they eventually will form zero-age
main-sequence OB stars which will rapidly heat, ionize, and disrupt
their surroundings.

We thus expect an evolutionary sequence for IRDCs as follows. The
youngest IRDCs will be isolated IR dark clouds, unassociated with any
bright, extended mid-IR or radio continuum emission. As protostars
form, the IRDC cores will be bright, compact far-IR sources but the
IRDC itself will remain dark. In the later stages, the IRDC will have
formed OB stars which will illuminate and ionize significant portions
of the cloud. Such IRDCs will contain very bright, extended mid-IR
emission and compact \hii\, regions. Eventually, the IRDC will lose
its `darkness' and become an IR bright star forming region.

We suggest that the degree to which IRDCs are associated with mid-IR
continuum emission sources is precisely due to this evolutionary
sequence. IRDCs with no associated mid-IR emission sources such as
\irdcfifteen\, (Fig.~\ref{mm-fig1}[a]) and \irdcthirty\,
(Fig.~\ref{mm-fig4}[a]) represent the very earliest stage before
high-mass protostars have formed. The intermediate stage might be
represented by \irdcfortythree\,
(Fig.~\ref{mm-fig5}[b]). \cite{Rathborne05} demonstrate that this IRDC
is the site of recent, active high-mass star formation. Examples of
the later evolutionary stage are \irdctwentysix\,
(Fig.~\ref{mm-fig3}[e]) and \irdceighteen\,
(Fig.~\ref{mm-fig6}[c]). Toward both of these IRDCs we see bright,
extended mid-IR emission. These bright sources appear to be \hii\,
regions breaking out of the dense cloud.

\subsection{IRDC Cores: Precursors to Hot Cores}

Whereas cluster formation takes place over size scales of several
parsecs and involves $\sim$10$^{3}$\,\Msun\, of molecular gas,
individual stars form from smaller, less massive sub-structures.
Individual low-mass stars form within Bok globules from
pre-protostellar cores with characteristic sizes of $\sim$0.05\,pc,
masses of 0.5--5\,\Msun, and densities of
$\sim$10$^{5}$--10$^{6}$\cmc\, (see Table~\ref{properties}). High-mass
stars should form from similar cold, compact cores, but with higher
masses and densities. However, the earliest well characterized stage
of high-mass star formation is the hot core phase, well after the
formation of the central protostar.

Just as IRDCs have similar sizes and masses compared to
cluster-forming clumps, the IRDC cores also share some of the
properties of hot cores that are associated with individual high-mass
stars. Both the IRDC cores ($\sim$ 0.3 pc) and hot cores ($<$ 0.1 pc)
are compact. In our sample, IRDC cores have a median mass of
120\,\Msun. We find that 67\% have masses in the range
30--300\,\Msun\, (see Fig.~\ref{histogram}), which is comparable to
the measured mass for hot cores $\sim$ 100--300\,\Msun\,
(\citealp{Garay99}).  Their densities are also comparable; IRDC cores
in our sample range from 10$^{3}$--10$^{7}$\,\cmc\, while hot cores,
have slightly larger densities ranging from 10$^{5}$--10$^{8}$\,\cmc\,
(see Table~\ref{properties}).

These similarities suggest that IRDC cores may be the cold precursors
to hot cores and the high-mass analogs of pre-protostellar cores
within Bok globules.  If true, then hot cores collapse from IRDC
cores, and therefore we should expect the IRDC cores to be somewhat
larger and less dense.  Indeed, the observations show that the IRDC
cores are factors of several larger and factors of $\sim$ 30 less
dense than hot cores.

The mass of the star that forms within an IRDC core depends on
fragmentation and the molecular core's star formation efficiency. In
the simplest scenario, a single core forms a single
star. Unfortunately, the core star formation efficiency is not well
known. \cite{Lada97} find efficiencies ranging from 10-30\% in
cluster-forming clumps. If we adopt a value of $\sim$ 20\% as a
typical core star formation efficiency, the majority of our cores will
form stars with masses ranging from 6 to 60\,\Msun\, corresponding to
OB stars. We conclude therefore that IRDC cores are the precursors to
high-mass protostars.  In fact, \cite{Rathborne05} find three
high-mass protostars associated with IRDC cores in
\irdcfortythree. 

We cannot yet exclude, however, the possibility that some IRDC cores
may instead further fragment to form several low- to intermediate-mass
stars. Indeed, the most massive ($>$500\,\Msun) cores with compact
($<$0.5 pc) sizes will likely further fragment into several
protostellar objects. Because the central protostar is directly
revealed by its millimeter through mid-IR emission, the number of
detected millimeter to mid-IR sources within these cores will trace
the degree of fragmentation, and more importantly, reveal the
luminosities of the individual protostars.  Thus, higher angular
resolution studies, in the millimeter/submillimeter and mid-IR, are
necessary to distinguish between single and multiple protostars. In
addition, such observations would also reveal the luminosities and,
hence, masses of the individual protostars.

Because high-mass IRDC cores are brighter at millimeter wavelengths
than low-mass cores, the lack of detected low-mass cores may result
from our limited sensitivity. Because the number of cores as a
function of mass N(M) is well characterized by a power law for masses
$>$100\,\Msun, one might expect the number of low-mass cores to follow
this trend. It is possible therefore that low-mass cores exist in
large numbers within IRDCs but are simply fainter than our detection
limit. Toward the nearby clouds, we do detect a few low-mass cores
($\sim$10\,\Msun); it is likely these will only give rise to low-mass
stars.

\subsection{Implications for the Role of IRDCs in Galactic Star Formation}

The discovery of a large number of IRDCs and cores has important
implications for their role in global star formation in the Galaxy. If
IRDCs spawn star clusters, and stars form predominantly in clusters
\citep{Lada03}, then every star in the Galaxy may very well form
within molecular clumps with identical temperatures, sizes, masses,
column densities, and volume densities as IRDCs. These clumps would
only appear as an IRDC if they happened to be nearby, with a strong
mid-IR background.

A very rough estimate of the star formation rate within IRDC-like
molecular clumps can be made as follows.  The total mass in the form
of IRDCs, $M_{IRDC,tot}$, can be estimated using

\begin{equation}
M_{IRDC,tot} = N_{IRDC} <M_{IRDC}> \frac{1}{f_{detect}}
\end{equation}

\noindent where $N_{IRDC}$ is the total number of known IRDCs in the Galaxy,
$<M_{IRDC}>$ is the characteristic mass of an individual IRDC, and
$f_{detect}$ is the fraction of all IRDCs that can be detected as
extinction features. Although our sample comes from the \MSX\,
database, new and future surveys will no doubt improve our
understanding of IRDCs due to improved sensitivity and angular
resolution. For example, the \spitzer\, 3--8\,\um\, Galactic Legacy
Infrared Mid-Plane Survey Extraordinaire (GLIMPSE;
\citealp{Benjamin03}), improves the angular resolution by an order of
magnitude over \MSX\, and therefore reveals new IRDCs.
\MSX, however, does detect the largest, most massive IRDCs that presumably
dominate Galactic star-formation. 

From the \MSX\, database, roughly 10,000 IRDCs have been identified
\citep{Simon-catalog}. If the IRDCs in our sample are typical, then we
can choose their average mass as a characteristic mass for IRDCs, and
so $<M_{IRDC}> \sim$ 3,000\,\Msun.  The total number of IRDCs in the
Galaxy is not yet known. Because of strong observational bias only a
fraction, $f_{detect}$, of all IRDCs are actually detected as
extinction features.  Toward the inner Galaxy, IRDCs at the far
kinematic distance will lie behind most of the diffuse IR emission and
would be impossible to detect as extinction features.  From
geometrical arguments, sources within the solar circle at the near
kinematic distance occupy 1/3 of the area of sources at the far
kinematic distance. Since we preferentially detect IRDCs at the near
kinematic distance we estimate $f_{detect} \sim 1/3$.  With these
estimates, the total Galactic mass contained in IRDCs is $\sim
10^{8}$\,\Msun, or about 5\% of the total molecular gas content of the
Milky Way.

The rate of star formation $\dot{M}$ occurring within IRDCs can now be estimated using

\begin{equation}
\dot{M}_{SF} = \frac{\eta M_{IRDC,tot}}{\tau_{SF}}
\end{equation}

\noindent where $\eta$ is the star formation efficiency, and $\tau_{SF}$ is the timescale
for star formation to occur within an IRDC.  If we choose an
efficiency of $\eta \sim 0.2$ (typical for cluster forming clumps) and
a conservative star formation timescale of $10^{7}$ yr, we arrive at a
total star formation within IRDCs of $\dot{M} \sim 2$ \Msun\,
yr$^{-1}$.  This star formation rate is quite close to estimates of
the global star formation of the Galaxy, $\sim$ 3--5 \Msun\, yr$^{-1}$
\citep{Prantzos95}.  While this estimate is only approximate, it does
support the plausibility of the idea that IRDC-like molecular clumps
are the birthplaces of all stars.

\subsection{Comparison to the Stellar IMF}

If stars form within IRDC cores, then it is also interesting to
compare the mass spectrum of IRDC cores with the stellar initial mass
function (IMF). If these high-mass cores spawn single high-mass stars
with a constant star formation efficiency, one might expect the power
law index $\alpha$ to match that of the IMF for high-mass
stars. Indeed, recent studies suggest that the stellar initial mass
function is determined by the fragmentation of a cloud into cores,
well before any stars have formed
\citep{Motte98,Testi98,Blitz99,Williams00,Churchwell02}.

Above $M\sim100$\,\Msun, we find the IRDC core mass spectrum is
approximately a power law with an index $\alpha$ of $2.1 \pm
0.4$. Within the errors, this power-law index is the same as that for
the stellar IMF for high-mass stars ($\alpha \sim 2.35$;
\citealp{Salpeter55}) and also to that of the clump mass spectrum for
molecular clouds ($\alpha \sim$ 1.8; \citealp{Kramer98,Simon01}).

\section{Conclusions}

We used MAMBO II on the IRAM 30 m telescope to image the 1.2 mm
continuum emission toward a sample of 38 IRDCs with known kinematic
distance. We find 1.2~mm continuum emission to be associated with each
of the IRDCs observed. The morphology of the 1.2 mm continuum emission
matches the mid-IR extinction features extremely well; some IRDCs are
extended and filamentary, while others are compact and smooth. Every
IRDC contains at least one compact ($\leq$ 0.5 pc) core, with most
IRDCs containing multiple cores.

IRDCs have masses in the range 120 to 16,000\,\Msun\, (with a median
mass of 940\,\Msun) and, in some cases, are associated with bright
mid-IR emission sources.  We suggest that the degree to which IRDCs
are associated with bright mid-IR emission sources is related to their
evolutionary phase; IRDCs with no associated mid-IR emission sources
are in the earliest phase, while IRDCs with several bright mid-IR
emission sources are in the later stages. The colder temperatures,
fragmented sub-structure, and similar sizes and masses of IRDCs and
cluster-forming molecular clumps, support our idea that IRDCs may be
the precursors to cluster-forming molecular clumps and the high-mass
analogs to Bok globules.

Toward the 38 IRDCs, we detected a total of 140 millimeter cores
unassociated with \MSX\, 8\,\um\, emission.  The core masses range
from 10 to 2,100 \Msun\, and have a median mass of 120\,\Msun. We find
that IRDC cores have similar sizes, masses, and densities to hot cores
associated with individual high-mass stars, the only obvious
difference being that IRDC cores are colder. Assuming typical star
formation efficiencies ($\sim$ 20\%), the simplest interpretation is
that the majority of the cores will form high-mass stars in the range
6 to 60\,\Msun. We suggest therefore that IRDC cores are in an earlier
evolutionary phase than hot cores and that they are the higher-mass
analog to the pre-protostellar cores within Bok globules.

Given the limited angular resolution and sensitivity of the 1.2 mm
continuum images, we cannot yet exclude the possibility that many of
the cores will further fragment to form low- to intermediate-mass
stars or that many low-mass cores exist within IRDCs but are below our
detection limit. However, an estimate of the star formation rate
within IRDC-like molecular clumps ($\sim$ 2\,\Msun\,yr$^{-1}$) is
comparable to the global star formation rate in the Galaxy, which
supports the idea that such molecular clumps may be the birthplaces of
all stars. We also find the slope of the IRDC core mass spectrum
($\alpha$ $\sim$ 2.1 $\pm$ 0.4) is similar to that of the stellar IMF.

Future high-resolution studies of IRDCs may help solve several
outstanding problems in cluster formation. For example, the location
of cores of various masses within IRDCs will test the idea that the
observed stellar mass segregation (the tendency for higher mass stars
to be located nearer to the center of a cluster) in evolved clusters
results from an earlier mass segregation of the proto-stellar cores
\citep{Gouliermis04}. In addition, emission from the extended regions
of IRDCs will trace the initial conditions and cloud density structure
just prior to fragmentation. IRDCs, therefore, represent excellent
laboratories to study the core mass spectrum, its relation to the
stellar initial mass function, and the observed mass segregation
within star clusters.

\acknowledgments

The authors gratefully acknowledge funding support through NASA grant
NNG04GGC92G. We thank Axel Weiss for help with the IRAM
observations. The data have been reduced with the mapping software
package developed by Robert Zylka. This software uses the
``restoring'' algorithm of \cite{Emerson79}, the ``converting''
algorithm of C. Salter (``Continuum Mapping with the NRAO 12-m
Telescope'' user's manual), and partly the NOD2 and the GILDAS
libraries.  IRAM is supported by INSU/CNRS (France), MPG (Germany) and
IGN (Spain).  This research made use of data products from the {\it
Midcourse Space Experiment}. Processing of the data was funded by the
Ballistic Missile Defense Organization with additional support from
NASA Office of Space Science.  This research has also made use of the
NASA/IPAC Infrared Science Archive, which is operated by the Jet
Propulsion Laboratory, California Institute of Technology, under
contract with the National Aeronautics and Space Administration. We
thank the anonymous referee for suggesting several clarifications and
augmentations that have greatly improved the paper.


\clearpage 

\clearpage 

\clearpage 

\clearpage 

\clearpage 

\clearpage 

\clearpage


\begin{table}
\caption{\label{properties}Comparison of molecular clump and core properties}
\begin{tabular}{lc|ccc|ccc}
\tableline
\tableline
\multicolumn{1}{c}{Properties} 
                   && Bok         & Cluster-forming    &  IRDCs            &\multicolumn{3}{c}{Cores}\\
                   && globules    &  clumps            &                   & pre-protostellar &  Hot      &IRDC\\
\tableline          
Radius (pc)       & & 0.1--2      &   0.5--1           & 1--3              &$\sim$0.05   & $<$0.1 & 0.02--0.8 \\
Mass (\Msun)      & & 1--10$^{2}$ & 10$^{2}$--10$^{3}$ & 10$^{2}$--10$^{4}$& 0.5--5 & $\sim$10$^{2}$ & 10--10$^{3}$\\
Density (\cmc)&&10$^{3}$--10$^{4}$&10$^{4}$--10$^{7}$&10$^{2}$--10$^{4}$&10$^{5}$--10$^{6}$&10$^{5}$--10$^{8}$& 10$^{3}$--10$^{7}$\\
Temperature (K)   & & 10--20      &   50--200          & 10-20             & 10     & 50--250    & 15--30 \\
\tableline
References        & & 1,2,3         & 4,5            & 6               & 7,8  & 9,10,11     & 12,13\\
\tableline
\end{tabular}
\tablerefs{(1) \cite{Leung85}; (2) \cite{Clemens88}; (3) \cite{Bourke95}; (4) \cite{Lada03}; (5) \cite{Motte03}; (6) this work; (7) \cite{Meyers83}; (8) \cite{Ward-Thompson94}; (9) \cite{Garay99}; (10) \cite{Kurtz00}; (11) \cite{Churchwell02}; (12) this work; (13) \cite{Rathborne05}.}
\end{table}

\clearpage
\begin{table}
\scriptsize
\caption{Properties of the IRDCs\label{table-clouds}}
\begin{tabular}{ccccccrrr}
\tableline
\tableline
{Name} & \multicolumn{2}{c}{Coordinates}   & {\vlsr}   & {$\Delta V$}  & {$D$} &  \multicolumn{2}{c}{1.2 mm Flux}  &  \multicolumn{1}{c}{Mass} \\
       & {$\alpha$}     & {$\delta$}  &  &  &   & \multicolumn{1}{c}{Peak} &  \multicolumn{1}{c}{Integrated} & \\
{MSXDC}   &{(J2000)} &{(J2000)}  & {(\kms)} & {(\kms)} & {(kpc)} & \multicolumn{1}{c}{(mJy)}     &  \multicolumn{1}{c}{(Jy)}  & \multicolumn{1}{c}{\Msun}\\
\tableline
 G015.05+00.07   &   18  17  41.8   &   -15  49  38.7   &  30.9  &  6.8  &  3.2  &  115  &  0.71  &  158  \\  
 G015.31$-$00.16 &   18  18  51.4   &   -15  43  27.7   &  31.1  &  3.3  &  3.2  &  41   &  1.67  &  370  \\  
 G018.82$-$00.28 &   18  26  19.8   &   -12  41  29.7   &  65.8  &  5.5  &  4.8  &  80   &  6.36  &  3168  \\ 
 G019.27+00.07   &   18  25  54.0   &   -12  04  56.3   &  26.2  &  4.5  &  2.4  &  150  &  3.25  &  405  \\  
 G022.35+00.41   &   18  30  24.7   &   -09  10  47.4   &  60.5  &  2.2  &  4.3  &  349  &  3.71  &  1483  \\ 
 G022.73+00.11   &   18  32  13.4   &   -09  00  58.5   &  77.8  &  7.0  &  5.1  &  38   &  0.72  &  405  \\  
 G023.60+00.00   &   18  34  21.0   &   -08  17  31.1   &  53.9  &  5.8  &  3.9  &  272  &  8.83  &  2903  \\ 
 G024.08+00.04   &   18  35  02.9   &   -07  46  11.8   &  52.5  &  5.8  &  3.8  &  68   &  3.04  &  949  \\ 
 G024.33+00.11   &   18  35  16.3   &   -07  36  09.3   &  52.9  &  6.5  &  3.8  &  1199 &  8.34  &  2603  \\ 
 G024.60+00.08   &   18  35  39.4   &   -07  18  50.9   &  51.7  &  3.9  &  3.7  &  263  &  4.22  &  1249  \\  
 G025.04$-$00.20 &   18  37  26.6   &   -07  09  14.2   &  46.9  &  4.2  &  3.4  &  92   &  6.07  &  1517  \\  
 G027.75+00.16   &   18  41  20.4   &   -04  32  25.1   &  79.1  &  6.9  &  5.3  &  75   &  0.99  &  599  \\  
 G027.94$-$00.47 &   18  44  01.5   &   -04  38  41.6   &  45.7  &  3.1  &  3.2  &  103  &  1.55  &  343  \\  
 G027.97$-$00.42 &   18  43  52.8   &   -04  36  13.6   &  45.9  &  3.5  &  3.2  &  143  &  1.61  &  356  \\  
 G028.04$-$00.46 &   18  44  10.3   &   -04  33  37.1   &  45.5  &  3.0  &  3.2  &  84   &  0.64  &  142  \\   
 G028.08+00.07   &   18  42  20.4   &   -04  16  50.3   &  75.8  &  5.4  &  4.9  &  63   &  0.72  &  374  \\  
 G028.10$-$00.45 &   18  44  13.6   &   -04  29  47.4   &  46.2  &  2.5  &  3.2  &  33   &  1.80  &  398  \\  
 G028.23$-$00.19 &   18  43  31.3   &   -04  13  18.4   &  79.6  &  6.5  &  5.1  &  83   &  2.80  &  1574  \\  
 G028.28$-$00.34 &   18  44  11.0   &   -04  17  39.2   &  47.4  &  4.5  &  3.3  &  64   &  6.19  &  1457  \\ 
 G028.37+00.07   &   18  42  50.6   &   -04  03  30.4   &  78.6  &  8.3  &  5.0  &  199  &  29.41 &  15895  \\  
 G028.53$-$00.25 &   18  44  17.1   &   -03  59  36.6   &  87.0  &  4.7  &  5.7  &  227  &  15.96 &  11210  \\ 
 G028.67+00.13   &   18  43  07.9   &   -03  44  28.5   &  79.5  &  8.0  &  5.1  &  70   &  1.67  &  939  \\  
 G030.14$-$00.06 &   18  46  36.4   &   -02  31  22.1   &  86.7  &  4.0  &  5.5  &  53   &  1.22  &  798  \\  
 G030.57$-$00.23 &   18  48  02.0   &   -02  12  40.1   &  86.2  &  4.0  &  5.5  &  55   &  2.47  &  1615  \\  
 G030.97$-$00.14 &   18  48  24.2   &   -01  48  24.7   &  78.8  &  6.1  &  5.1  &  27   &  5.08  &  2856  \\ 
 G031.02$-$00.10 &   18  48  22.2   &   -01  45  03.2   &  78.0  &  5.9  &  5.0  &  45   &  0.30  &  157  \\  
 G031.97+00.07   &   18  49  33.7   &   -00  47  48.3   &  96.7  &  6.2  &  6.9  &  311  &  16.32 &  14797  \\ 
 G033.69$-$00.01 &   18  52  57.6   &    00  42  58.9   &  105.9 &  6.1  &  7.1  &  205  &  12.14 &  13230  \\  
 G034.43+00.24   &   18  53  18.9   &    01  26  38.6   &  57.1  &  5.8  &  3.7  &  2228 &  36.88 &  10914  \\ 
 G034.77$-$00.55 &   18  56  48.8   &    01  23  21.0   &  43.5  &  5.0  &  2.9  &  59   &  4.85  &  882  \\  
 G035.39$-$00.33 &   18  57  09.0   &    02  07  45.7   &  44.7  &  3.6  &  2.9  &  76   &  8.40  &  1527  \\  
 G035.59$-$00.24 &   18  57  11.7   &    02  15  54.7   &  44.7  &  4.3  &  2.9  &  57   &  0.67  &  121  \\ 
 G036.67$-$00.11 &   18  58  40.4   &    03  16  17.7   &  54.4  &  3.6  &  3.6  &  61   &  3.92  &  1098  \\  
 G038.95$-$00.47 &   19  04  08.3   &    05  08  49.3   &  41.6  &  3.1  &  2.7  &  119  &  11.37 &  1792  \\ 
 G048.65$-$00.29 &   19  21  45.3   &    13  49  21.7   &  34.0  &  2.8  &  2.5  &  71   &  6.79  &  917  \\  
 G053.11+00.05   &   19  29  18.1   &    17  54  33.0   &  22.0  &  2.2  &  1.8  &  81   &  7.41  &  519  \\  
 G053.25+00.04   &   19  29  37.5   &    18  01  23.5   &  23.9  &  2.1  &  1.9  &  96   &  4.36  &  340  \\  
 G053.31+00.00   &   19  29  54.5   &    18  03  30.0   &  24.0  &  2.7  &  2.0  &  71   &  4.54  &  393  \\  
\tableline
References       & 1                 & 1                & 2      & 2     & 2     &       &        &    \\
\tableline
\end{tabular}
\tablerefs{(1) Simon et al. 2006a; (2) Simon et al. 2006b}
\end{table}


\clearpage
\begin{deluxetable}{cllccccccc}
\tabletypesize{\scriptsize}
\tablecaption{Properties of the IRDC cores \label{table-cores}}
\tablewidth{0pt}
\tablehead{
\colhead{} &\colhead{Core} & \multicolumn{2}{c}{Coordinates}  & \multicolumn{2}{c}{1.2 mm Flux} &\multicolumn{2}{c}{FWHM Diameter}  & 
\colhead{Mass}   & \colhead{Comment\tablenotemark{a}} \\
\colhead{} & \colhead{}      & \colhead{$\alpha$}      & \colhead{$\delta$}      & \colhead{Peak} & \colhead{Integrated} & \colhead{Angular} & 
\colhead{Physical} & \colhead{} &\colhead{} \\
\colhead{} &\colhead{}       & \colhead{(J2000)} & \colhead{(J2000)}               &\colhead{(mJy)}     & \colhead{(Jy)}             & \colhead{($\arcsec$)} & \colhead{(pc)}       & \colhead{(\Msun)} & \colhead{}}
\startdata
\multicolumn{4}{l}{MSXDC  G015.05+00.07}       &      &        &      &        &      &    \\   %
  & MM1   &   18  17  50.4   &   -15  53  38   &  115 &  0.47  &  24  &  0.32  &  105 &    \\   
  & MM2   &   18  17  40.0   &   -15  48  55   &  50  &  0.37  &  32  &  0.46  &  83  & (b)\\   
  & MM3   &   18  17  42.4   &   -15  47  03   &  35  &  0.10  &  22  &  0.28  &  22  &    \\   
  & MM4   &   18  17  32.0   &   -15  46  35   &  35  &  0.19  &  28  &  0.39  &  43  &    \\   
  & MM5   &   18  17  40.2   &   -15  49  47   &  30  &  0.13  &  24  &  0.33  &  29  &    \\   
\multicolumn{4}{l}{ MSXDC  G015.31-00.16 }     &      &        &      &        &      &    \\   %
  & MM1   &   18  18  56.4   &   -15  45  00   &  41  &  0.17  &  24  &  0.33  &  38  &    \\   
  & MM2   &   18  18  50.4   &   -15  43  19   &  39  &  0.46  &  40  &  0.59  &  101 &    \\   
  & MM3   &   18  18  45.3   &   -15  41  58   &  37  &  0.74  &  52  &  0.79  &  164 &    \\   
  & MM4   &   18  18  48.0   &   -15  44  22   &  23  &  0.08  &  24  &  0.32  &  18  &    \\   
  & MM5   &   18  18  49.1   &   -15  42  47   &  21  &  0.14  &  40  &  0.59  &  32  &    \\   
\multicolumn{4}{l}{ MSXDC  G018.82-00.28  }    &      &        &      &        &      &    \\   %
  & MM1   &   18  25  56.1   &   -12  42  48   &  459 &  1.78  &  23  &  0.46  &  887 &  em \\   
  & MM2   &   18  26  23.4   &   -12  39  37   &  170 &  0.49  &  20  &  0.37  &  242 &  em \\   
  & MM3   &   18  25  52.6   &   -12  44  37   &  118 &  0.47  &  25  &  0.52  &  233 &  em \\   
  & MM4   &   18  26  15.5   &   -12  41  32   &  80  &  0.53  &  31  &  0.66  &  263 &    \\   
  & MM5   &   18  26  21.0   &   -12  41  11   &  55  &  0.34  &  30  &  0.64  &  170 &    \\   
  & MM6   &   18  26  18.4   &   -12  41  15   &  45  &  0.41  &  35  &  0.77  &  205 &    \\   
 \multicolumn{4}{l}{MSXDC  G019.27+00.07 }     &      &        &      &        &      &    \\   %
  & MM1   &   18  25  58.5   &   -12  03  59   &  150 &  0.91  &  29  &  0.30  &  113 &    \\   
  & MM2   &   18  25  52.6   &   -12  04  48   &  89  &  0.92  &  37  &  0.41  &  114 &    \\   
 \multicolumn{4}{l}{MSXDC  G022.35+00.41 }     &      &        &      &        &      &    \\   %
  & MM1   &   18  30  24.4   &   -09  10  34   &  349 &  0.63  &  16  &  0.22  &  253 & (a)\\   
  & MM2   &   18  30  24.2   &   -09  12  44   &  67  &  0.54  &  33  &  0.64  &  215 &   em\\   
  & MM3   &   18  30  38.1   &   -09  12  44   &  53  &  0.29  &  28  &  0.52  &  114 &    \\   
 \multicolumn{4}{l}{ MSXDC  G022.73+00.11}     &      &        &      &        &      &    \\   %
  & MM1   &   18  32  13.0   &   -09  00  50   &  38  &  0.27  &  31  &  0.70  &  149 &    \\   
 \multicolumn{4}{l}{ MSXDC  G023.60+00.00}     &      &        &      &        &      &    \\   %
  & MM1   &   18  34  11.6   &   -08  19  06   &  375 &  1.11  &  20  &  0.31  &  365 &   em\\   
  & MM2   &   18  34  21.1   &   -08  18  07   &  272 &  0.71  &  19  &  0.29  &  233 &    \\   
  & MM3   &   18  34  10.0   &   -08  18  28   &  81  &  0.81  &  41  &  0.74  &  266 &   em\\   
  & MM4   &   18  34  23.0   &   -08  18  21   &  68  &  1.07  &  47  &  0.85  &  350 &    \\   
  & MM5   &   18  34  09.5   &   -08  18  00   &  65  &  0.30  &  26  &  0.43  &  98  &   em\\   
  & MM6   &   18  34  18.2   &   -08  18  52   &  61  &  0.29  &  27  &  0.45  &  96  &    \\   
  & MM7   &   18  34  21.1   &   -08  17  11   &  58  &  0.87  &  47  &  0.85  &  286 &    \\   
  & MM8   &   18  34  24.9   &   -08  17  14   &  40  &  0.14  &  22  &  0.34  &  45  &    \\   
  & MM9   &   18  34  22.5   &   -08  16  04   &  39  &  0.24  &  31  &  0.53  &  80  &    \\   
 \multicolumn{4}{l}{ MSXDC  G024.08+00.04}     &      &        &      &        &      &    \\   %
  & MM1   &   18  34  57.0   &   -7  43  26    &  219 &  0.63  &  20  &  0.29  &  196 &   em\\   
  & MM2   &   18  34  51.1   &   -7  45  32    &  68  &  0.64  &  40  &  0.70  &  201 &    \\   
  & MM3   &   18  35  2.2    &   -7  45  25    &  50  &  0.40  &  33  &  0.56  &  124 &    \\   
  & MM4   &   18  35  2.6    &   -7  45  56    &  48  &  0.37  &  32  &  0.55  &  115 &    \\   
  & MM5   &   18  35  7.4    &   -7  45  46    &  40  &  0.28  &  32  &  0.54  &  87  &    \\   
 \multicolumn{4}{l}{MSXDC  G024.33+00.11}      &      &        &      &        &      &    \\   %
  & MM1   &   18  35  07.9   &   -07  35  04   & 1199 &  2.03  &  15  &  0.18  &  1759& IRAS 18324-0737   \\   
  & MM2   &   18  35  34.5   &   -07  37  28   &  117 &  0.39  &  25  &  0.40  &  123 &  em \\   
  & MM3   &   18  35  27.9   &   -07  36  18   &  96  &  0.41  &  25  &  0.40  &  126 &   em\\   
  & MM4   &   18  35  19.4   &   -07  37  17   &  90  &  1.43  &  48  &  0.85  &  446 &    \\   
  & MM5   &   18  35  33.8   &   -07  36  42   &  79  &  0.62  &  34  &  0.58  &  192 &   em\\   
  & MM6   &   18  35  07.7   &   -07  34  33   &  77  &  0.94  &  41  &  0.72  &  293 &    \\   
  & MM7   &   18  35  09.8   &   -07  39  48   &  77  &  0.30  &  24  &  0.38  &  93  &    \\   
  & MM8   &   18  35  23.4   &   -07  37  21   &  72  &  0.98  &  44  &  0.78  &  305 &    \\   
  & MM9   &   18  35  26.5   &   -07  36  56   &  66  &  0.88  &  46  &  0.82  &  273 &   em\\   
  & MM10  &   18  35  27.9   &   -07  35  32   &  64  &  0.31  &  26  &  0.42  &  97  &    \\   
  & MM11  &   18  35  05.1   &   -07  35  58   &  48  &  0.88  &  56  &  1.00  &  275 &    \\   
  \multicolumn{4}{l}{MSXDC  G024.60+00.08}     &      &        &      &        &      &    \\   %
  & MM1   &   18  35  40.2   &   -07  18  37   &  279 &  0.65  &  18  &  0.25  &  192 &    \\   
  & MM2   &   18  35  35.7   &   -07  18  09   &  230 &  0.53  &  18  &  0.42  &  483 &    \\   
  & MM3   &   18  35  41.1   &   -07  18  30   &  71  &  0.12  &  16  &  0.20  &  35  &    \\   
  & MM4   &   18  35  39.3   &   -07  18  51   &  30  &  0.09  &  23  &  0.36  &  27  &    \\        
 \multicolumn{4}{l}{MSXDC  G025.04-00.20 }     &      &        &      &        &      &    \\   %
  & MM1   &   18  38  10.2   &   -07  02  34   &  203 &  1.11  &  28  &  0.42  &  276 &   em\\   
  & MM2   &   18  38  17.7   &   -07  02  51   &  92  &  0.42  &  28  &  0.41  &  104 &    \\   
  & MM3   &   18  38  10.2   &   -07  02  44   &  90  &  0.19  &  18  &  0.22  &  47  &    \\   
  & MM4   &   18  38  13.7   &   -07  03  12   &  82  &  1.23  &  46  &  0.73  &  307 &    \\   
  & MM5   &   18  38  12.0   &   -07  02  44   &  73  &  0.20  &  21  &  0.29  &  50  &    \\   
 \multicolumn{4}{l}{MSXDC  G027.75+00.16 }     &      &        &      &        &      &    \\   %
  & MM1   &   18  41  19.9   &   -04  32  20   &  75  &  0.53  &  31  &  0.73  &  320 & (a)\\   
  & MM2   &   18  41  33.0   &   -04  33  44   &  46  &  0.22  &  27  &  0.63  &  133 &    \\   
  & MM3   &   18  41  16.8   &   -04  31  55   &  36  &  0.46  &  42  &  1.04  &  281 &    \\   
  & MM4   &   18  41  30.4   &   -04  30  00   &  29  &  0.20  &  32  &  0.77  &  123 &    \\   
  & MM5   &   18  41  23.6   &   -04  30  42   &  28  &  0.15  &  28  &  0.64  &  92  &    \\   
 \multicolumn{4}{l}{ MSXDC  G027.94-00.47 }    &      &        &      &        &      &    \\   %
  & MM1   &   18  44  03.6   &   -04  38  00   &  103 &  0.32  &  21  &  0.26  &  70  &    \\   
  & MM2   &   18  44  03.1   &   -04  37  25   &  40  &  0.19  &  27  &  0.37  &  42  &    \\   
  \multicolumn{4}{l}{MSXDC  G027.97-00.42  }   &      &        &      &        &      &    \\   %
  & MM1   &   18  43  52.8   &   -04  36  13   &  143 &  0.26  &  16  &  0.16  &  57  & (a)\\   
  & MM2   &   18  43  58.0   &   -04  34  24   &  103 &  0.28  &  19  &  0.23  &  62  &   (b)\\   
  & MM3   &   18  43  54.9   &   -04  36  08   &  41  &  0.27  &  18  &  0.20  &  45  &    \\   
 \multicolumn{4}{l}{ MSXDC  G028.04-00.46 }    &      &        &      &        &      &    \\   %
  & MM1   &   18  44  08.5   &   -04  33  22   &  87  &  0.36  &  24  &  0.32  &  80  &   em\\   
 \multicolumn{4}{l}{ MSXDC  G028.08+00.07  }   &      &        &      &        &      &    \\   %
  & MM1   &   18  42  20.3   &   -04  16  42   &  63  &  0.72  &  39  &  0.89  &  374 &    \\   
\multicolumn{4}{l}{MSXDC  G028.10-00.45  }     &      &        &      &        &      &    \\   %
  & MM1   &   18  44  12.9   &   -04  29  45   &  33  &  0.35  &  38  &  0.56  &  78  &    \\   
  & MM2   &   18  44  14.3   &   -04  29  48   &  18  &  0.09  &  28  &  0.40  &  19  &    \\   
\multicolumn{4}{l}{ MSXDC  G028.23-00.19}      &      &        &      &        &      &    \\   %
  & MM1   &   18  43  30.7   &   -04  13  12   &  83  &  1.25  &  51  &  1.22  &  705 & (a)\\   
  & MM2   &   18  43  29.0   &   -04  12  16   &  35  &  0.19  &  27  &  0.60  &  108 &   em\\   
  & MM3   &   18  43  30.0   &   -04  12  33   &  23  &  0.07  &  21  &  0.43  &  38  &    \\   
 \multicolumn{4}{l}{MSXDC  G028.28-00.34  }    &      &        &      &        &      &    \\   %
  & MM1   &   18  44  15.0   &   -04  17  54   &  317 &  1.74  &  27  &  0.39  &  411 &   em\\   
  & MM2   &   18  44  21.3   &   -04  17  37   &  165 &  1.36  &  37  &  0.55  &  321 &   em\\   
  & MM3   &   18  44  13.4   &   -04  18  05   &  139 &  0.29  &  17  &  0.20  &  68  &   em\\   
  & MM4   &   18  44  11.4   &   -04  17  22   &  64  &  0.59  &  37  &  0.55  &  139 &    \\   
\multicolumn{4}{l}{ MSXDC  G028.37+00.07 }     &      &        &      &        &      &    \\   %
  & MM1   &   18  42  52.1   &   -03  59  45   &  470 &  2.12  &  26  &  0.55  &  1148&   em\\   
  & MM2   &   18  42  37.6   &   -04  02  05   &  277 &  1.00  &  22  &  0.46  &  542 &   em\\   
  & MM3   &   18  43  03.1   &   -04  06  24   &  248 &  0.89  &  23  &  0.47  &  482 &   em\\   
  & MM4   &   18  42  50.7   &   -04  03  15   &  199 &  0.61  &  21  &  0.43  &  329 &    \\   
  & MM5   &   18  42  26.8   &   -04  01  30   &  151 &  0.33  &  17  &  0.30  &  177 &   em\\   
  & MM6   &   18  42  49.0   &   -04  02  23   &  145 &  0.43  &  20  &  0.40  &  232 &    \\   
  & MM7   &   18  42  56.3   &   -04  07  31   &  138 &  0.56  &  24  &  0.51  &  304 &   em\\   
  & MM8   &   18  42  49.7   &   -04  09  54   &  115 &  0.77  &  31  &  0.70  &  414 &   em\\   
  & MM9   &   18  42  46.7   &   -04  04  08   &  97  &  0.73  &  34  &  0.77  &  397 &    \\   
  & MM10  &   18  42  54.0   &   -04  02  30   &  88  &  0.66  &  34  &  0.79  &  358 &    \\   
  & MM11  &   18  42  42.7   &   -04  01  44   &  81  &  0.83  &  38  &  0.88  &  447 &    \\   
  & MM12  &   18  43  09.9   &   -04  06  52   &  71  &  0.88  &  42  &  0.97  &  476 &    \\   
  & MM13  &   18  42  41.8   &   -03  57  08   &  66  &  0.86  &  42  &  0.98  &  463 &    \\   
  & MM14  &   18  42  52.6   &   -04  02  44   &  62  &  0.07  &  12  &  0.08  &  36  &    \\   
  & MM15  &   18  42  32.4   &   -04  01  16   &  59  &  0.25  &  26  &  0.55  &  134 &   em\\   
  & MM16  &   18  42  40.2   &   -04  00  23   &  58  &  0.83  &  44  &  1.03  &  447 &    \\   
  & MM17  &   18  43  00.0   &   -04  01  34   &  56  &  0.37  &  30  &  0.67  &  198 &    \\   
  & MM18  &   18  43  12.9   &   -04  01  16   &  54  &  0.30  &  34  &  0.76  &  162 &    \\   
 \multicolumn{4}{l}{MSXDC  G028.53-00.25  }    &      &        &      &        &      &    \\   %
  & MM1   &   18  44  18.0   &     -03  59  34 &  227 &  1.66  &  33  &  0.84  &  1165& (a)\\   
  & MM2   &   18  44  15.7   &     -03  59  41 &  129 &  3.01  &  56  &  1.51  &  2115&    \\   
  & MM3   &   18  44  16.0   &     -04  00  48 &  126 &  2.91  &  56  &  1.51  &  2044&    \\   
  & MM4   &   18  44  18.6   &     -04  00  05 &  95  &  1.14  &  44  &  1.16  &  800 &    \\   
  & MM5   &   18  44  17.0   &     -04  02  04 &  95  &  0.33  &  23  &  0.53  &  234 &    \\   
  & MM6   &   18  44  17.8   &     -04  00  05 &  71  &  0.17  &  18  &  0.38  &  119 &    \\   
  & MM7   &   18  44  23.7   &     -04  02  09 &  63  &  0.72  &  45  &  1.19  &  508 & (b)\\   
  & MM8   &   18  44  22.0   &     -04  01  35 &  61  &  0.26  &  27  &  0.66  &  185 &    \\   
  & MM9   &   18  44  19.3   &     -03  58  05 &  58  &  0.24  &  26  &  0.63  &  166 &    \\   
  & MM10  &   18  44  18.5   &     -03  58  43 &  57  &  0.62  &  41  &  1.07  &  433 &    \\   
\multicolumn{4}{l}{ MSXDC  G028.67+00.13 }     &      &        &      &        &      &    \\   %
  & MM1   &   18  43  03.1   &   -03  41  41   &  86  &  0.25  &  20  &  0.40  &  143 & em  \\   
  & MM2   &   18  43  07.1   &   -03  44  01   &  70  &  0.70  &  38  &  0.90  &  394 &    \\   
  & MM3   &   18  42  58.2   &   -03  48  20   &  60  &  0.22  &  22  &  0.46  &  122 & em  \\   
  & MM4   &   18  43  13.2   &   -03  41  03   &  37  &  0.21  &  29  &  0.64  &  116 &    \\   
  & MM5   &   18  43  10.1   &   -03  45  08   &  32  &  0.18  &  30  &  0.67  &  103 &    \\   
  & MM6   &   18  43  12.2   &   -03  45  39   &  31  &  0.20  &  31  &  0.71  &  113 &    \\   
  & MM7   &   18  43  06.9   &   -03  45  11   &  27  &  0.16  &  30  &  0.67  &  87  &    \\   
\multicolumn{4}{l}{ MSXDC  G030.14-00.06 }     &      &        &      &        &      &    \\   %
  & MM1   &   18  46  35.7   &   -02  31  03   &  53  &  0.90  &  51  &  1.32  &  597 &    \\   
  & MM2   &   18  46  31.7   &   -02  32  41   &  36  &  0.13  &  23  &  0.53  &  83  &    \\   
\multicolumn{4}{l}{ MSXDC  G030.57-00.23 }     &      &        &      &        &      &    \\   %
  & MM1   &   18  48  00.0   &   -02  07  20   &  297 &  0.60  &  17  &  0.33  &  419 & em  \\   
  & MM2   &   18  47  58.7   &   -02  15  20   &  143 &  0.30  &  17  &  0.33  &  198 & em  \\   
  & MM3   &   18  47  54.5   &   -02  11  15   &  55  &  0.39  &  31  &  0.77  &  257 &    \\   
  & MM4   &   18  48  01.8   &   -02  12  35   &  37  &  0.46  &  45  &  1.15  &  300 & (a)\\   
  & MM5   &   18  47  59.9   &   -02  11  01   &  36  &  0.22  &  32  &  0.78  &  144 &    \\   
  & MM6   &   18  47  59.4   &   -02  13  13   &  33  &  0.16  &  31  &  0.77  &  104 &    \\   
 \multicolumn{4}{l}{MSXDC  G030.97-00.14 }     &      &        &      &        &      &    \\   %
  & MM1   &   18  48  21.6   &   -01  48  27   &  114 &  0.74  &  32  &  0.73  &  417 & em  \\   
  & MM2   &   18  48  22.0   &   -01  47  42   &  27  &  0.41  &  47  &  1.13  &  231 &    \\   
\multicolumn{4}{l}{ MSXDC  G031.02-00.10  }    &      &        &      &        &      &    \\   %
  & MM1   &   18  48  20.7   &   -01  44  48   &  130 &  0.65  &  24  &  0.50  &  349 & em  \\   
 \multicolumn{4}{l}{MSXDC  G031.97+00.07   }   &      &        &      &        &      &    \\   %
  & MM1   &   18  49  36.3   &   -00  45  45   &  913 &  1.84  &  17  &  0.40  &  1890& em; IRAS  18470-0049   \\   
  & MM2   &   18  49  36.0   &   -00  46  16   &  311 &  0.90  &  20  &  0.55  &  929 &    \\   
  & MM3   &   18  49  32.3   &   -00  47  02   &  187 &  1.19  &  30  &  0.91  &  1222&    \\   
  & MM4   &   18  49  33.0   &   -00  47  33   &  117 &  0.83  &  32  &  0.98  &  852 &    \\   
  & MM5   &   18  49  21.9   &   -00  50  35   &  96  &  0.17  &  16  &  0.35  &  178 & (b)\\   
  & MM6   &   18  49  35.0   &   -00  46  44   &  96  &  1.15  &  41  &  1.32  &  1181&    \\   
  & MM7   &   18  49  28.4   &   -00  48  54   &  80  &  0.28  &  23  &  0.67  &  291 &    \\   
  & MM8   &   18  49  29.1   &   -00  48  12   &  64  &  0.48  &  33  &  1.03  &  493 &    \\   
  & MM9   &   18  49  31.6   &   -00  46  30   &  47  &  0.15  &  21  &  0.57  &  151 & em  \\   
  \multicolumn{4}{l}{ MSXDC  G033.69-00.01 }   &      &        &      &        &      &    \\   %
  & MM1   &   18  52  58.8   &   00  42  37    &  205 &  1.04  &  27  &  0.82  &  1135&    \\   
  & MM2   &   18  52  49.9   &   00  37  57    &  115 &  1.23  &  42  &  1.37  &  1342& em  \\   
  & MM3   &   18  52  50.8   &   00  36  43    &  81  &  0.26  &  22  &  0.63  &  288 & em  \\   
  & MM4   &   18  52  56.4   &   00  43  08    &  78  &  0.75  &  37  &  1.19  &  820 &    \\   
  & MM5   &   18  52  47.8   &   00  36  47    &  56  &  0.22  &  24  &  0.73  &  243 & em; IRAS 18502+00.33   \\   
  & MM6   &   18  52  48.7   &   00  35  58    &  47  &  0.36  &  32  &  1.03  &  395 &    \\   
  & MM7   &   18  52  58.1   &   00  44  08    &  43  &  0.59  &  47  &  1.57  &  641 &    \\   
  & MM8   &   18  52  53.9   &   00  41  16    &  41  &  0.54  &  44  &  1.46  &  588 &    \\   
  & MM9   &   18  52  58.1   &   00  41  20    &  40  &  0.11  &  19  &  0.52  &  119 &    \\   
  & MM10  &   18  52  52.7   &   00  38  35    &  40  &  0.10  &  19  &  0.52  &  114 &    \\   
  & MM11  &   18  52  56.2   &   00  41  48    &  31  &  0.09  &  21  &  0.60  &  100 &    \\   
 \multicolumn{4}{l}{ MSXDC  G034.43+00.24}     &      &        &      &        &      &    \\   %
  & MM1   &   18  53  18.0   &   01  25  24    &  2228&  4.01  &  16  &  0.19  &  1187&    \\   
  & MM2   &   18  53  18.6   &   01  24  40    &  964 &  4.33  &  26  &  0.42  &  1284&   em; IRAS 18507+0121   \\   
  & MM3   &   18  53  20.4   &   01  28  23    &  244 &  1.02  &  24  &  0.38  &  301 &    \\   
  & MM4   &   18  53  19.0   &   01  24  08    &  221 &  0.86  &  24  &  0.38  &  253 &    \\   
  & MM5   &   18  53  19.8   &   01  23  30    &  122 &  2.24  &  51  &  0.89  &  664 &    \\   
  & MM6   &   18  53  18.6   &   01  27  48    &  57  &  0.43  &  37  &  0.62  &  126 &    \\   
  & MM7   &   18  53  18.3   &   01  27  13    &  55  &  0.29  &  28  &  0.46  &  87  &    \\   
  & MM8   &   18  53  16.4   &   01  26  20    &  51  &  0.36  &  31  &  0.52  &  108 &    \\   
  & MM9   &   18  53  18.4   &   01  28  14    &  50  &  0.53  &  39  &  0.67  &  157 &    \\   
 \multicolumn{4}{l}{ MSXDC  G034.77-00.55 }    &      &        &      &        &      &    \\   %
  & MM1   &   18  56  48.2   &   01  18  47    &  232 &  0.91  &  23  &  0.28  &  166 &  em \\   
  & MM2   &   18  56  50.3   &   01  23  16    &  59  &  1.04  &  49  &  0.66  &  188 &    \\   
  & MM3   &   18  56  44.7   &   01  20  42    &  44  &  0.08  &  16  &  0.15  &  14  &    \\   
  & MM4   &   18  56  48.9   &   01  23  34    &  42  &  0.38  &  36  &  0.47  &  70  &    \\   
 \multicolumn{4}{l}{ MSXDC  G035.39-00.33  }   &      &        &      &        &      &    \\   %
  & MM1   &   18  56  41.2   &   02  09  52    &  128 &  0.42  &  21  &  0.25  &  76  &   em\\   
  & MM2   &   18  56  59.2   &   02  04  53    &  114 &  0.25  &  17  &  0.18  &  45  &   em\\   
  & MM3   &   18  57  05.3   &   02  06  29    &  107 &  0.44  &  24  &  0.30  &  79  &   em\\   
  & MM4   &   18  57  06.7   &   02  08  23    &  76  &  0.59  &  34  &  0.45  &  108 &    \\   
  & MM5   &   18  57  08.8   &   02  08  09    &  69  &  0.65  &  37  &  0.49  &  118 &    \\   
  & MM6   &   18  57  08.4   &   02  09  09    &  56  &  0.39  &  37  &  0.49  &  71  &    \\   
  & MM7   &   18  57  08.1   &   02  10  50    &  55  &  0.53  &  39  &  0.52  &  96  & (c)\\   
  & MM8   &   18  57  07.0   &   02  08  54    &  52  &  0.33  &  29  &  0.37  &  59  &    \\   
  & MM9   &   18  57  11.2   &   02  07  27    &  50  &  0.23  &  27  &  0.34  &  42  &    \\   
  \multicolumn{4}{l}{MSXDC  G035.59-00.24  }   &      &        &      &        &      &    \\   %
  & MM1   &   18  57  02.3   &   02  17  04    &  69  &  0.36  &  28  &  0.36  &  65  &   em\\   
  & MM2   &   18  57  07.4   &   02  16  14    &  57  &  0.26  &  25  &  0.31  &  48  &    \\   
  & MM3   &   18  57  11.6   &   02  16  08    &  44  &  0.23  &  27  &  0.34  &  41  &    \\   
 \multicolumn{4}{l}{ MSXDC  G036.67-00.11}     &      &        &      &        &      &    \\   %
  & MM1   &   18  58  39.6   &   03  16  16    &  61  &  0.17  &  21  &  0.31  &  49  &(a)\\   
  & MM2   &   18  58  35.6   &   03  15  06    &  54  &  0.21  &  24  &  0.36  &  58  &    \\   
 \multicolumn{4}{l}{ MSXDC  G038.95-0.47 }     &      &        &      &        &      &    \\   %
  & MM1   &   19  04  07.4   &   05  08  48    &  119 &  0.74  &  29  &  0.35  &  117 &(a) \\   
  & MM2   &   19  04  03.4   &   05  07  56    &  103 &  0.46  &  25  &  0.28  &  73  &   em\\   
  & MM3   &   19  04  07.4   &   05  09  44    &  65  &  0.07  &  12  &  0.04  &  11  &    \\   
  & MM4   &   19  04  00.6   &   05  09  06    &  53  &  0.31  &  33  &  0.40  &  48  &   em\\   
 \multicolumn{4}{l}{ MSXDC  G048.65-00.29 }    &      &        &      &        &      &    \\   %
  & MM1   &   19  21  49.7   &   13  49  30    &  71  &  0.39  &  29  &  0.32  &  52  &    \\   
  & MM2   &   19  21  47.6   &   13  49  22    &  57  &  0.29  &  26  &  0.28  &  39  &    \\   
 \multicolumn{4}{l}{ MSXDC  G053.11+00.05  }   &      &        &      &        &      &    \\   %
  & MM1   &   19  29  17.2   &   17  56  21    &  536 &  1.77  &  21  &  0.15  &  124 &   em\\   
  & MM2   &   19  29  20.2   &   17  57  06    &  81  &  0.63  &  33  &  0.27  &  44  &    \\   
  & MM3   &   19  29  00.6   &   17  55  11    &  70  &  0.18  &  19  &  0.13  &  12  &   em\\   
  & MM4   &   19  29  20.4   &   17  55  04    &  52  &  0.64  &  44  &  0.37  &  45  &    \\   
  & MM5   &   19  29  26.3   &   17  54  53    &  48  &  0.19  &  24  &  0.18  &  13  &   em\\   
 \multicolumn{4}{l}{ MSXDC  G053.25+00.04 }    &      &        &      &        &      &    \\   %
  & MM1   &   19  29  39.0   &   18  01  42    &  246 &  0.31  &  13  &  0.06  &  24  &   em\\   
  & MM2   &   19  29  33.0   &   18  01  00    &  97  &  0.95  &  38  &  0.33  &  74  &   em\\   
  & MM3   &   19  29  44.0   &   17  58  47    &  96  &  0.13  &  14  &  0.07  &  10  &    \\   
  & MM4   &   19  29  34.5   &   18  01  39    &  71  &  0.15  &  17  &  0.12  &  12  &    \\   
  & MM5   &   19  29  39.4   &   17  58  40    &  52  &  0.25  &  30  &  0.25  &  19  &    \\   
  & MM6   &   19  29  31.5   &   17  59  50    &  47  &  0.22  &  26  &  0.21  &  17  &   em\\   
  \multicolumn{4}{l}{MSXDC  G053.31+00.00 }    &      &        &      &        &      &    \\   %
  & MM1   &   19  29  50.0   &   18  05  07    &  71  &  0.36  &  29  &  0.26  &  31  &    \\   
  & MM2   &   19  29  42.1   &   18  03  57    &  60  &  0.38  &  30  &  0.26  &  33  &    \\   
  & MM3   &   19  29  49.7   &   18  04  39    &  21  &  0.09  &  29  &  0.25  &  8   &    \\   
\enddata

\tablenotetext{a}{Includes: ``em'' marks cores that are associated with 8\,\um\, emission in the \MSX\, images; letters
in parentheses refer to the corresponding cores from the IRDC catalog of \cite{Simon-catalog}; the names
of IRAS sources associated with the millimeter cores are also listed here.}
\end{deluxetable}

\end{document}